\documentclass[pra,twocolumn,showpacs,amsmath,amssymb,amsfonts]{revtex4}
\usepackage{bm}
\usepackage{graphicx}
\usepackage{subfigure}
\usepackage{array}
\usepackage{color}

\def\Ham{\mathcal{H}}
\def\nop{\hat{n}}
\newcommand{\mbf}[1]{{\textbf #1}}
\newcommand{\mrm}{\mathrm}
\newcommand{\bepsilon}{\mbox{\boldmath$\epsilon$}}
\newcommand{\bra}[1]{\langle #1|}
\newcommand{\ket}[1]{|#1\rangle}

\newcommand{\aver}[1]{\left\langle #1 \right\rangle}

\begin{document}

\title{Thermal effects in light scattering from ultracold bosons in an optical lattice}

\author{Kazimierz {\L}akomy}
\affiliation{Institute of Theoretical Physics, Faculty of Physics,
  University of Warsaw, Ho\.{z}a 69, 00-681 Warsaw, Poland}
\author{Zbigniew Idziaszek}
\affiliation{Institute of Theoretical Physics, Faculty of Physics,
  University of Warsaw, Ho\.{z}a 69, 00-681 Warsaw, Poland}
\author{Marek Trippenbach}
\affiliation{Institute of Theoretical Physics, Faculty of Physics,
  University of Warsaw, Ho\.{z}a 69, 00-681 Warsaw, Poland}
\date{\today}

\begin{abstract}
  We study the scattering of a weak and far-detuned light from a system of ultracold
  bosons in 1D and 3D optical lattices. We show the connection between angular
  distributions of the scattered light and statistical properties of a Bose gas
  in a periodic potential. The angular patterns are determined by the Fourier
  transform of the second-order correlation function, and thus they can be
  used to retrieve information on particle number fluctuations and correlations.
  We consider superfluid and Mott insulator phases of the Bose gas in a lattice,
  and we analyze in detail how the scattering depends on the system
  dimensionality, temperature and atom-atom interactions.
\end{abstract}
\maketitle

\section{Introduction}

Experimental realization of Bose-Einstein condensates (BEC) in ultracold
trapped gases opened up a rapidly expanding field of studies of quantum-degenerate
systems \cite{Dalfovo:1999,Giorgini:2008,Bloch:2008}. Among the others,
statistical properties of the condensate, such as its fluctuations and
correlations, have attracted a wide attention. While the theory on this subject 
is well developed (see e.g. \cite{Idziaszek3} and references therein), there are 
only few experiments that address this issue. To date, only the fluctuations of the 
total number of atoms in a condensed gas have been measured \cite{Raizen}, and the 
subpoissonian scaling has been observed. The second order correlation functions, 
that are directly connected to the condensate atom number fluctuations, have been 
investigated experimentally in the collisions of metastable helium condensates 
\cite{Westbrook} and in the expanding rubidium condensate \cite{Schmiedmayer}.

One of the potential tools to measure the statistics of quantum-degenerate gases is 
based on atom-light interactions.This possibility has been noticed already some 
time ago, and has been proposed for a detection of the Bose condensed phase
\cite{Lewenstein:1993,Javanainen:1995,Saito:1999,Moore:1999}, superfluidity in 
Fermi gases \cite{Zhang:1999,Ruostekoski:1999,Torma:2000,Wong:2000} and, quite 
recently, for a detection of quantum phases in ultracold gases in optical lattices 
\cite{Mekhov:2007,Mekhov:2007a,Chen:2007,Eckert:2007}. The optical imaging techniques 
have already been used to measure coherence properties of BEC in the Raman superradiant 
scattering \cite{sadler:2007} and the second order correlation functions 
\cite{Schmiedmayer}. 

The light scattered from a quantum gas carries information on atoms statistics, 
and thus can be used to measure the condensate fluctuations \cite{IdziaszekScatt}. 
In the case of BEC in a trap the profile of the scattered light is dominated by a 
component, which depends on the mean occupation number of the condensate. 
In order to detect a much weaker component resulting from fluctuations, 
one has to resort to the variance of the number of scattered photons, 
which can be difficult to measure.

The situation changes, however, in the presence of a periodic potential.
In this case the dominating classical component exhibits interference
pattern characteristic for the Bragg scattering, and the quantum component
can be measured at the angles corresponding to destructive interference,
where the large classical component vanishes. This property has been first
noticed by Mekhov {\it et al.} \cite{Mekhov:2007}, and these authors have
proposed a method of probing the statistics of an ultracold gas in a lattice
\cite{Mekhov:2007b}, based on the relatively strong coupling between atoms
and light modes of a cavity. In this case one should be able to perform
non-demolition measurement allowing to distinguish between superfluid
and Mott-insulator (MI) quantum phases at temperature $T=0$.

In this paper we study a less complex situation of measuring a quantum gas
statistics based on the far-off-resonance light scattering from a Bose gas
in a lattice, focusing on the effects of statistics at finite temperatures.
In order to avoid atom losses and suppress a possibility of perturbing the
quantum state by the probing laser, we assume that the probing light is
sufficiently weak and far-detuned. We show that the mean number of photons
detected at some special angles, carries enough information not only to
distinguish between different thermodynamic phases of the gas but also to
directly measure the effects of the on-site atom statistics driven by
quantum and thermal fluctuations. Hence, it allows to verify the validity
of some well-grounded literature approaches, such as the Bogoliubov method,
to describe higher order correlation functions in an interacting Bose gas.

Our paper is organized as follows. In section \ref{sec:atom-light} we develop
a model for scattering of light from ultracold atoms, showing that the number
of scattered photons is directly related to the second-order correlation function.
In section \ref{sec:opt-latt} we tailor our model to the external potential created
by an optical lattice. The scattering from atoms in one dimensional (1D) lattice
is considered in section \ref{sec:oneDimTzeroScatt}, where for simplicity we focus
only on the zero-temperature statistics discussing the effects of different approximations.
The finite temperature statistics of a Bose gas in a lattice is analyzed in
section \ref{sec:statProperties}. Section \ref{sec:3D} investigates scattering
from atoms in three-dimensional (3D) optical lattice at finite temperatures. We
conclude in section \ref{sec:conclusions}, and finally two appendixes present
technical details related to the influence of non-local Franck-Condon coefficients
(Appendix A) and optimal configuration of a probing laser and a photon detector
in the 3D lattice case (Appendix B).

\section{Interaction of light with many atoms}
\label{sec:atom-light}

In this section we consider a general problem of light scattering from a gas
of bosons in an arbitrary external potential. We assume that the trapped atoms
are illuminated with a weak, and far-detuned laser light. The angularly resolved
scattered light is measured by detectors in the far-field region. The full
Hamiltonian of the system consists of the following parts:
\begin{equation}
  \Ham=\Ham_{a} + \Ham_{f} + \Ham_{al} + \Ham_{af}
\end{equation}
where $\Ham_{a}$ is the atomic Hamiltonian, $\Ham_{f}$ represents vacuum modes
of the electromagnetic field (EM), $\Ham_{al}$ describes interaction of atoms
with the laser light
and $\Ham_{af}$ interaction of atoms with vacuum modes.

The atomic Hamiltonian can be split into two parts
\begin{equation}
  \Ham_{a}=\Ham_{0} + \Ham_{int}
\end{equation}
where $\Ham_{0}$ describes the system of two-level atoms in the second-quantization
formalism \cite{Lewenstein}
\begin{equation}
  \Ham_{0}=\sum\limits_{\mbf{n}} \hbar \omega^{g}_{\mbf{n}} g^{\dagger}_{\mbf{n}} 
  g_{\mbf{n}} + \sum\limits_{\mbf{m}} \hbar \left( \omega_{\mbf{m}}^{e}
    +\omega_{0}\right) e^{\dagger}_{\mbf{m}} e_{\mbf{m}},
\end{equation}
and the part including the atom-atom interactions reads
\begin{equation}
  \Ham_{int} = \frac{1}{2} \sum\limits_{\mbf{n}, \mbf{m}, \mbf{p}, \mbf{q}} 
  U_{\mbf{n} \mbf{m} \mbf{p} \mbf{q}}
  \,g_{\mbf{n}}^{\dagger} g_{\mbf{m}}^{\dagger} g_{\mbf{p}} g_{\mbf{q}}.
\end{equation}
Here, $g_{\mbf{n}}$ ($g^{\dagger}_{\mbf{n}}$) is the annihilation (creation)
operator of an atom in the ground electronic state and state $\mbf{n}$ of the
center-of-mass (COM) motion, and $e_{\mbf{m}}$ ($e^{\dagger}_{\mbf{m}}$) is
the annihilation (creation) operator of an atom in an electronic excited
state and state $\mbf{m}$ of COM motion. The operators obey the standard
bosonic commutation relations:
$[g_{\mbf{n}},g_{\mbf{m}}^{\dagger}]=\delta_{\mbf{n},\mbf{m}}$ and
$[e_{\mbf{n}},e_{\mbf{m}}^{\dagger}]=\delta_{\mbf{n},\mbf{m}}$.
The corresponding eigenenergies of atom COM motion are denoted by
$\hbar \omega^{g}_{\mbf{n}}$ and $\hbar \omega_{\mbf{m}}^{e}$ for atoms in the
ground and excited electronic states, respectively. The matrix elements
$U_{\mbf{n} \mbf{m} \mbf{p} \mbf{q}}$ of the interaction Hamiltonian read
\begin{align}
  U_{\mbf{n} \mbf{m} \mbf{p} \mbf{q}} \equiv \frac{4 \pi a_s \hbar^2}{m} 
  \int d^{3}r \phi^{\ast}_{\mbf{n}}(\mbf{r})
  \phi^{\ast}_{\mbf{m}}(\mbf{r}) \phi_{\mbf{p}}(\mbf{r}) \phi_{\mbf{q}}(\mbf{r})
  \label{eqn:Hamint}
\end{align}
where we model short-range interactions through a contact potential with $s$-wave
scattering length $a_s$ and a mass of the atom $m$. We neglect ground-excited
and excited-excited atom interactions, assuming that for a weak and far-detuned
probing light excited atoms constitute only a small fraction of the whole sample.

The Hamiltonian of the EM field takes a standard form:
\begin{equation}
  \Ham_{f}=\sum\limits_{\lambda} \int 
  d^{3}k \ \hbar \omega_{\mbf{k}} a^{\dagger}_{\mbf{k} \lambda} a_{\mbf{k} \lambda}
  \label{eqn:HamEM}
\end{equation}
with $a_{\mbf{k} \lambda}$ ($a^{\dagger}_{\mbf{k} \lambda}$) being an annihilation
(creation) operator of a photon with a wave vector $\mbf{k}$ and a polarization 
$\lambda$.

The interaction of atoms with a laser beam is described as follows:
\begin{equation}
  \Ham_{al}=\frac{\hbar \Omega}{2} \sum\limits_{\mbf{n},\mbf{m}} \bra{\mbf{n},g} 
  u_{\mbf{k}_L}(\mbf{r}) \ket{\mbf{m},e} e^{\imath \omega_{L} t} g^{\dagger}_{\mbf{n}} 
  e_{\mbf{m}} + h.c.
  \label{eqn:Hamal}
\end{equation}
where we treat the macroscopically occupied laser mode classically. Here,
$u_{\mbf{k}_L}(\mbf{r})$ characterizes a laser mode with a wave vector $\mbf{k}_L$, $\omega_{L}$
is the laser frequency, $\Omega$ is a Rabi frequency of the atomic transition,
and the Franck-Condon coefficients $\bra{\mbf{n},g} u_{\mbf{k}_L}(\mbf{r}) \ket{\mbf{m},e}$
describe a transition amplitude between COM motion states $\mbf{n}$ and $\mbf{m}$
of the atoms in the ground and excited electronic states, respectively. Typically,
for a single probing laser, we have $u_{\mbf{k}_L}(\mbf{r}) = e^{\imath \mbf{k}_L \mbf{r}}$
(running wave), and for the two counter-propagating probing beams
$u_{\mbf{k}_L}(\mbf{r}) = \cos\left( \mbf{k}_L \mbf{r}\right)$ (standing wave).
In general $u_{\mbf{k}_L}(\mbf{r})$ can also represent the modes of an optical 
cavity \cite{Mekhov:2007}.

The part of the Hamiltonian that describes coupling of atoms with quantized EM 
field is given by:
\begin{align}
  \Ham_{af} & = \imath \sum\limits_{\lambda} \int d^{3}k \, \hbar c_{\mbf{k}\lambda} 
  a^{\dagger}_{\mbf{k} \lambda} \sum\limits_{\mbf{n}, \mbf{m}} \bra{\mbf{n},g} 
  u_{\mbf{k}}(\mbf{r}) \ket{\mbf{m},e} g^{\dagger}_{\mbf{n}} e_{\mbf{m}} \nonumber \\
  & \phantom{=} + h.c.
\end{align}
in which $c_{\mbf{k}\lambda} = \sqrt{\omega_{\mbf{k}}/(16 \pi^3 \epsilon_{0} \hbar)} 
\left(\textbf{d} \cdot \bepsilon_{\mbf{k} \lambda}\right)$, $\mbf{d}$ is a dipole 
moment of the atomic transition, $u_{\mbf{k}}(\mbf{r})$ is a mode function of the 
EM field with a wavevector $\mbf{k}$ and frequency $\omega_{\mbf{k}}$, and 
$\bepsilon_{\textbf{k}\lambda}$ is a unit vector perpendicular to $\mbf{k}$ 
describing the mode of light with a polarization $\lambda$.

We solve the quantum equations of motion in the Heisenberg picture under the
following approximation: i) we assume that the atomic operators are driven only
by the dominating laser mode of the EM field, neglecting the back action of
atoms on the laser mode, ii) the quantum dynamics of the vacuum modes is
determined by the evolution of atomic operators, ignoring the back-action
of the vacuum modes, which is equivalent to neglecting the process of
spontaneous emission, iii) for the weak and far-detuned laser field we
perform adiabatic elimination of the weakly populated excited state. Our
approximations are analogous to those used in \cite{IdziaszekScatt}, with
the only difference that here we perform the adiabatic elimination of the 
excited state, instead of assuming short probing pulses. We carry out our 
derivation neglecting the interactions between atoms and we comment on the 
generalization to the interacting gas case at the end of this section. 

The equations of motion for the atomic operators in the interaction picture 
with respect to $\Ham_{0}$:~$\tilde{g}_\mbf{m}(t) = g_\mbf{m}(t) e^{-\imath \omega_\mbf{m}^g t}$
and $\tilde{e}_\mbf{n}(t) = e_\mbf{n}(t) e^{-\imath (\omega_\mbf{n}^e + \omega_0) t}$, read
\begin{align}
  \label{eq:MotionAtomic1}
  \frac{d \tilde{g}_\mbf{m}}{d \tau} & = - \imath \frac{\Omega}{2 \Delta}  \sum_{\mbf{n}} 
  \eta_{\mbf{n}\mbf{m}}^\ast (\mbf{k}_L) \exp\left[\imath \frac{\omega_\mbf{m}^g 
      - \omega_\mbf{n}^e + \Delta}{\Delta}\tau\right] \tilde{e}_\mbf{n}(\tau), \\
  \label{eq:MotionAtomic2}
  \frac{d \tilde{e}_\mbf{n}}{d \tau} & = - \imath \frac{\Omega}{2 \Delta}  \sum_{\mbf{m}} 
  \eta_{\mbf{n}\mbf{m}} (\mbf{k}_L) \exp\left[\imath \frac{\omega_\mbf{n}^e 
      - \omega_\mbf{m}^g - \Delta}{\Delta} \tau\right] \tilde{g}_\mbf{m}(\tau),
\end{align}
where $\eta_{\mbf{n}\mbf{m}} (\mbf{k}) = \bra{\mbf{n},e} u_{\mbf{k}}(\mbf{r}) \ket{\mbf{m},g}$, 
$\Delta = \omega_{L} - \omega_{0}$ and we have introduced rescaled time variable 
$\tau = \Delta t$. We solve Eqs.~\eqref{eq:MotionAtomic1} and \eqref{eq:MotionAtomic2} 
by applying the Laplace transformation $F^{\cal L}(s) 
= \int_0^\infty d\tau \, e^{-s \tau} F(\tau)$. The Laplace transformed equations take the form
\begin{align}
  \label{eq:Lapl1}
  s \tilde{g}_\mbf{m}^{\cal L}(s) - \tilde{g}_\mbf{m}(0)  & = \nonumber \\ 
  - \imath \frac{\Omega}{2 \Delta} \sum_{\mbf{n}} & \eta_{\mbf{n}\mbf{m}}^\ast (\mbf{k}_L) 
  \tilde{e}_\mbf{n}^{\cal L} (s) \left[s -\imath \frac{\omega_\mbf{m}^g 
      - \omega_\mbf{n}^e + \Delta}{\Delta} \right],\\
  \label{eq:Lapl2}
  s \tilde{e}_\mbf{n}^{\cal L}(s) - \tilde{e}_\mbf{n}(0) & = \nonumber \\ 
  - \imath \frac{\Omega}{2 \Delta} \sum_{\mbf{m}} & \eta_{\mbf{n}\mbf{m}} (\mbf{k}_L) 
  \tilde{g}_\mbf{m}^{\cal L} (s) \left[s -\imath \frac{\omega_\mbf{n}^e 
      - \omega_\mbf{m}^g + \Delta}{\Delta}\right].
\end{align}
For a far-detuned light the prefactor on the right-hand-side of the equations is small: 
$\Omega/\Delta  \ll 1$,  and, in principle, the equations can be solved by iterations 
in a perturbative manner. Here, however, we proceed with solving Eq.~\eqref{eq:Lapl2} 
for $\tilde{e}_\mbf{n}^{\cal L}(s)$ and then substituting the result into Eq.~\eqref{eq:Lapl1}. 
For a far-detuned light we apply $\Delta \gg \omega_\mbf{n}^e, \omega_\mbf{m}^g$ and 
we use the identity $\displaystyle \sum_{\mbf{n}} \eta_{\mbf{n}\mbf{m}}^\ast (\mbf{k}_L) \eta_{\mbf{n}\mbf{m}^\prime} 
(\mbf{k}_L) = \delta_{\mbf{m}\mbf{m}^\prime}$, which results in
\begin{equation}
  \label{eq:Lapl3}
  \tilde{g}^{\cal L}_\mbf{m} (s) \approx \frac{1}{s + \imath \frac{\Omega^2}{\Delta^2}}
  \left( \tilde{g}_\mbf{m}(0) - \imath\frac{\Omega}{2 \Delta} \sum_\mbf{n} 
    \eta_{\mbf{n}\mbf{m}}^\ast (\mbf{k}_L) \frac{\tilde{e}_\mbf{n}(0)}{s-\imath} \right).
\end{equation}
By substituting back this result into Eq.~\eqref{eq:Lapl2}, and performing the 
inverse Laplace transformation we obtain the following time-dependence of the 
atomic operators
\begin{align}
  \label{eq:sol1}
  \tilde{g}_\mbf{m}(t) & = \tilde{g}_\mbf{m}(0) e^{- \imath \omega_\mrm{AC} t} 
  + {\cal O}\left(\frac{\Omega}{\Delta}\right), \\
  \tilde{e}_\mbf{n}(t) & = \tilde{e}_\mbf{n}(0) \nonumber \\
  & \, + \frac{\Omega}{2 \Delta} \sum_\mbf{m}
  \eta_{\mbf{n}\mbf{m}}(\mbf{k}_L) \tilde{g}_\mbf{m}(0) \left[
    e^{\imath(\omega_\mbf{n}^e - \omega_\mbf{m}^g - \Delta - \omega_\mrm{AC}) t} -1 \right] \nonumber \\
  & \, + {\cal O}\left(\frac{\Omega^2}{\Delta^2}\right),
  \label{eq:sol2}
\end{align}
Here,  $\omega_\mrm{AC} = \frac{\Omega^2}{4 \Delta}$ denotes AC Stark shift of atomic 
levels in the field of the probing laser. In Eq.~\eqref{eq:sol1} we have not included 
terms of the order of $\Omega/\Delta$ , which are proportional to $\tilde{e}_\mbf{n} (0)$, 
since they do not give any contribution to the mean number of photons, assuming that 
there are no excited atoms at the beginning.

We substitute Eqs.~\eqref{eq:sol1} and \eqref{eq:sol2} into equation of motion of the 
E-M field operators $\tilde{a}_\mbf{k \lambda}(t) = a_\mbf{k \lambda}(t) e^{-\imath \omega_\mbf{k} t}$ 
in the interaction picture. In the lowest order in $\Omega/\Delta$ this yields
\begin{multline}
  \label{eq:sola}
  \tilde{a}_\mbf{k \lambda}(t) - \tilde{a}_\mbf{k \lambda}(0) = \\
  = c_{\mbf{k} \lambda} \frac{\Omega}{\Delta} \sum_{\mbf{n}\mbf{n}^\prime\mbf{m}}
  \eta_{\mbf{m}\mbf{n}^\prime}(\mbf{k}) \eta_{\mbf{m}\mbf{n}}^\ast(\mbf{k}_L) 
  \tilde{g}^\dagger_\mbf{n}(0) \tilde{g}_{\mbf{n}^\prime}(0) \\
  \times \frac{e^{\imath (\omega_\mbf{k} -\omega^L_{\mbf{n} \mbf{n}^\prime})t/2}}
  {\omega_\mbf{k} -\omega^L_{\mbf{n} \mbf{n}^\prime}} \sin 
  \left( \frac{\omega_\mbf{k} -\omega^L_{\mbf{n} \mbf{n}^\prime}}{2} t \right)
\end{multline}
where $\omega^L_{\mbf{n} \mbf{n}^\prime} \equiv \omega_L + \omega_{\mbf{n}^\prime}^g -\omega_\mbf{n}^g$. 
At $t \rightarrow \infty$ the sine term will produce a term proportional to the delta 
function, describing the energy conservation in the process of a single photon scattering: 
$\omega_\mbf{k} = \omega_{\mbf{k}_L} + \omega_{\mbf{n}^\prime}^g -\omega_\mbf{n}^g$. However, in our 
case we are interested in the total number of photons scattered into a given solid angle,
and not in the spectrum of the scattered light. Hence, we use the approximation
$\omega^L_{\mbf{n} \mbf{n}^\prime} \approx \omega_L$. This condition is also applicable in the 
physical systems where the natural linewidth $\Gamma$ associated with the atomic transition
is broader than frequencies of atom COM motion: $\Gamma \gg \omega_\mbf{n}^g$. 

Now, by using Eq.~\eqref{eq:sola} and approximation 
$\omega^L_{\mbf{n} \mbf{n}^\prime} \approx \omega_L$  we calculate the mean number 
of photons with a wavevector $\mbf{k}$ and a polarization $\lambda$ 
\begin{align}
  \label{MeanPhotonNumber}
  \!\!\left\langle a^{\dagger}_{\mbf{k} \lambda} \left(t\right) a_{\mbf{k} \lambda} \left(t\right) 
  \right\rangle = \frac{\Omega^2  c_{\mbf{k} \lambda}^2}{\Delta^2} \frac{\sin^{2} 
    \left( \left( \omega_{\mbf{k}} - \omega_{L} \right) t/2 \right) }{\left( \omega_{\mbf{k}} 
      - \omega_{L} \right)^2} F(\mbf{k}, \mbf{k}_{L}),
\end{align}
where the function $F(\mbf{k}, \mbf{k}_{L})$ is defined as follows:
\begin{align}
  F(\mbf{k}, \mbf{k}_{L}) \equiv & \sum_{\substack{ {\mbf{n}},{\mbf{n}}' \\ {\mbf{m}},{\mbf{m}}'}}
  \bra{\mbf{n}} u_{\mbf{k}}^{\ast}({\mbf{r}}) u_{\mbf{k}_L}({\mbf{r}}) \ket{\mbf{n}'} \bra{\mbf{m}}
  u_{\mbf{k}}(\mbf{r}) u_{\mbf{k}_L}^{\ast}(\mbf{r}) \ket{\mbf{m}'} \nonumber \\
  & \times \aver{g_{\mbf{n}}^{\dagger}(0) g_{\mbf{n}'}(0) g_{\mbf{m}}^{\dagger}(0) g_{\mbf{m}'}(0)}.
  \label{eqn:Fq}
\end{align}
Notice that, in Eq.~\eqref{eqn:Fq} all the
matrix elements are calculated between COM states of ground-state atoms
$|\mbf{n},g \rangle$ and to shorten the notation $|\mbf{n}\rangle \equiv |\mbf{n},g\rangle$.
In the particular case when the mode functions $u_{\mbf{k}}(\mbf{r})$
and $u_{\mbf{k}_{L}}(\mbf{r})$ are the plane waves, $F(\mbf{k}, \mbf{k}_{L})$
reduces to the Fourier transform of the second-order correlation function
in atomic field operators $\hat{\Psi}_{g}(\mbf{x})$ of the atoms in the electronic ground state
\begin{align}
  F(\mbf{q}) = \int \!\!d^{3}x \int \!\!d^{3}y\, e^{\imath \mbf{q}(\mbf{x}-\mbf{y})}
  \left\langle \hat{\Psi}_{g}^{\dagger}(\mbf{x}) \hat{\Psi}_{g}(\mbf{x}) 
    \hat{\Psi}_{g}^{\dagger}(\mbf{y}) \hat{\Psi}_{g}(\mbf{y}) \right\rangle
\end{align}
where $\mbf{q} = \mbf{k} - \mbf{k}_L$ is the wave vector of the momentum transfer.
In the rest of the paper we will use the $F(\mbf{q})$ function only. 

An analogous result is obtained when considering the scattering of neutrons from liquid 
helium \cite{vanHove}. In that case the number of scattered particles associated with 
the momentum transfer $\mbf{q}$ and the energy transfer to the system $\hbar \omega$ is 
described by the dynamic structure factor
\begin{multline}
  S(\mbf{q},\omega) \equiv
  \frac{1}{N} \int \!\!d^{3}x \!\! \int \!\!d^{3}y \,\, e^{\imath \mbf{q}(\mbf{x}-\mbf{y})} \\
  \times \langle \Psi_E|
  \hat{\rho}(\mbf{x}) \delta (H - E -\hbar \omega) \hat{\rho}(\mbf{x}) | \Psi_E \rangle
\end{multline}
where $\hat{\rho}(\mbf{x}) = \hat{\Psi}^{\dagger}(\mbf{x}) \hat{\Psi}(\mbf{x})$,
$H$ is the Hamiltonian of the system, and $|\Psi_E\rangle$ is an eigenstate with energy $E$. 
By integrating over energies of the scattered particles one obtains the static structure factor
\begin{equation}
  S(\mbf{q}) = \hbar \int_{-\infty}^{\infty} d\omega\, S(\mbf{q},\omega)
\end{equation}
which is equivalent to our function $F(\mbf{q})$ describing an amplitude of scattered 
photons integrated over photon frequencies \cite{rem1}. We will refer to $F(\mbf{q})$
as the structure function in the rest of the paper.

For evolution time $t$ much longer than the time scale determined by
the optical frequencies $\omega_L$, we can apply the
following identity
\begin{equation}
  \lim_{t \rightarrow \infty} 
  \frac{\sin^{2} \left( \left( \omega_{\mbf{k}} - \omega_{L} \right) t/2 \right)}
  {t \left( \omega_{\mbf{k}} - \omega_{L} \right)^2} 
  = \frac{\pi}{2} \,\delta \left( \omega_{\mbf{k}} - \omega_{L} \right),
  \label{lim}
\end{equation}
to show that for the weak and far-detuned laser the scattered light
described by Eq.~\eqref{MeanPhotonNumber} has spectrum centered around 
elastic component.

The total number of photons scattered into a solid angle $d\Omega$ is equal to
\begin{equation}
  \frac{d N_{ph}}{d \Omega} (\hat{\mbf{k}}) = \sum\limits_{\lambda} \int\!\! dk\; k^2
  \aver{a^{\dagger}_{\mbf{k} \lambda} \left(t\right) a_{\mbf{k} \lambda} \left(t\right)}
  \label{eqn:distrib}
\end{equation}
where $\hat{\mbf{k}}=\mbf{k}/|\mbf{k}|$ represents the direction of
measurement. Since, according to Eq.~\eqref{lim}, the number of photons
is proportional to pulse length as expected, it is more convenient to calculate
the number of photons scattered into $d\Omega$ per unit of time
\begin{align}
  \frac{d^2 N_{ph}}{d\Omega dt} (\mbf{k},\mbf{k}_L) & = \frac{\Omega^2 \omega_{L}^{3} d^2}
  {32 \pi^2 \Delta^2 \epsilon_{0} \hbar c^3} {\cal W}(\hat{\mbf{k}}) F(\mbf{k}-\mbf{k}_L) 
  \nonumber \\
  & = \left[ \frac{d^2 N_{ph}}{d\Omega dt} (\mbf{k},\mbf{k}_L) \right]_{
    \substack{\!\!\mrm{one}\\\mrm{atom}}} F(\mbf{q})
  \label{eqn:scaling}
\end{align}
where ${\cal W}(\hat{\mbf{k}})= \left( 1-( \bepsilon_{\textbf{d}} 
  \cdot\bepsilon_{\textbf{k}})^2 \right)$ is the dipole pattern of the 
emitted light and $\mbf{\bepsilon}_{\textbf{d}}$ is a unit vector in the 
direction of the dipole moment $\textbf{d}$ that is determined by a 
polarization of the probing laser. Eq.~\eqref{eqn:scaling}
shows that the angular distribution of the scattered light, apart from the
contribution from the dipole pattern, is determined only by $F(\mbf{q})$.
In addition, for a single atom $F(\mbf{q})~=~1 $ and thus all the information
about scattering from the system of $N$ atoms is contained in $F(\mbf{q})$.
Therefore, in the subsequent sections, we can focus solely on the properties
of $F(\mbf{q})$, keeping in mind that the remaining contribution is the same
as for the scattering from a single atom.

The generalization of our derivation to the case of interacting atoms can be 
performed in an analogy to the problem of neutron scattering from liquid helium 
\cite{vanHove,Pitaevskii}. If one applies the Born approximation, and eliminates 
adiabatically the excited state, one ends up with the result identical to the one 
presented here. A similar approach has been applied to the study of Raman scattering 
in the superradiant regime \cite{uys:2008}.

Finally we note that our perturbative treatment neglects the effects of
the momentum transfer resulting from the photon recoil in the process of
light scattering. We assume, however, that the scattered light is weak and
far-detuned, therefore we expect that the fraction of atoms which experience
the photon recoil is sufficiently small, such that the atom statistics is not
significantly affected. Moreover, in the presence of a tight trapping potential, 
such as a deep optical lattice, one finds that the scattering is recoilless
\cite{gajda:1996}, which requires the trap size smaller than a wave length of
the scattered light.

\section{Scattering from ultracold gas of bosons in an optical lattice}
\label{sec:opt-latt}

\begin{figure}[b]
  \includegraphics[scale=0.8]{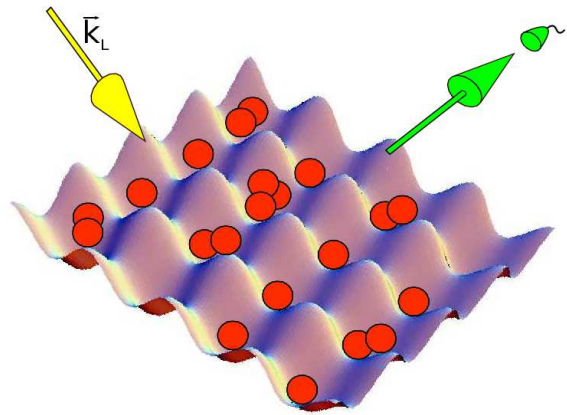}
  \caption{(Color online) Setup. An ultracold gas of bosons confined in an 
    optical lattice is illuminated with a probing laser beam (yellow arrow) 
    characterized by the wavevector $\mbf{k}_{L}$. The photons scattered into 
    a selected direction (green arrow) are collected by a detector.}
  \label{fig:setup}
\end{figure}

The setup we consider is schematically plotted in Fig.~\ref{fig:setup}. 
It consists of an ultracold gas of $N$ bosons confined in an optical cubic 
lattice of $M$ sites. We assume a homogeneous system with an equal average 
number of atoms $n = N/M$ in each site of a lattice. The periodic potential 
of the lattice reads \cite{Bloch:2008}:
\begin{equation}
  V_{p}(x,y,z) = V_{0} \left( \sin^2 k_{p} x + \sin^2 k_{p} y 
    + \sin^2 k_{p} z \right)
  \label{eqn:lattPotential}
\end{equation}
where $\mbf{k}_{p}$ is a wave vector of laser beams that are used to form the 
lattice and $V_{0}$ is the potential depth. The exact configuration of the 
probing beam and detectors will depend on a dimensionality of the lattice and 
will be discussed later. In order to use the results of the previous section, 
we need to specify a single-particle basis. In the case of atoms confined in 
an optical lattice it is convenient to choose the basis of Wannier functions 
$w_\mbf{m}(\mbf{r})$ that represent wave functions localized at single lattice 
sites $\mbf{m}$ and are linear combinations of Bloch states. In our approach we 
consider only excitations within the lowest Bloch band, so in a limit of deep 
optical lattices the Wannier functions describe only the ground state wave 
functions in local potential wells.

\subsection{Deep lattice regime}

For a deep optical lattice, the Wannier states are well localized within 
the sites of the lattice, and in equation (\ref{eqn:Fq}) we can restrict to 
optical transitions between the states localized at the same lattice sites: 
$\mbf{n} = \mbf{n}'$ and $\mbf{m} = \mbf{m}'$. In this case
$F(\mbf{q})$ simplifies to the following expression
\begin{align}
  F(\mbf{q}) &= \sum\limits_{\mbf{n},\mbf{m}} \bra{\mbf{n}} e^{\imath \mbf{q} \mbf{r}} 
  \ket{\mbf{n}} \bra{\mbf{m}} e^{-\imath \mbf{q} \mbf{r}} \ket{\mbf{m}}
  \aver{g_{\mbf{n}}^{\dagger} g_{\mbf{n}} g_{\mbf{m}}^{\dagger} g_{\mbf{m}}} \nonumber \\
  &= \left| f_{\textbf{0},\textbf{0}} (\mbf{q}) \right|^2 \sum\limits_{\mbf{n},\mbf{m}} 
  e^{ \imath \mbf{q} \left( \mbf{r}_{\mbf{n}} - \mbf{r}_{\mbf{m}} \right) } 
  \aver{n_{\mbf{n}} n_{\mbf{m}}}
  \label{eqn:FdeepLatt}
\end{align}
where $n_{\mbf{m}} \equiv g_{\mbf{m}}^{\dagger} g_{\mbf{m}}$ and
\begin{equation}
  f_{\mbf{n}, \mbf{m}}(\mbf{q}) \equiv \bra{\mbf{n}} e^{\imath \mbf{q} \mbf{r}} 
  \ket{\mbf{m}} = \int d^{3}r \,w^{\ast}_{\mbf{n}}(\mbf{r}) e^{\imath \mbf{q} \mbf{r}} 
  w_{\mbf{m}}(\mbf{r}).
\end{equation}
In analogy to the scattering of light into an optical cavity 
\cite{Mekhov:2007b}, we can define the classical part $F^{clas}(\mbf{q})$ 
and the quantum part $F^{quant}(\mbf{q})$ of the function $F(\mbf{q})$
\begin{align}
  F^{clas}(\mbf{q}) & \equiv n^2 \left| f_{\textbf{0},\textbf{0}} (\mbf{q}) \right|^2 
  \left| \sum\limits_{\mbf{m}} e^{\imath \mbf{q} \mbf{r}_{\mbf{m}}} \right|^{2},
  \label{eqn:FClasDef} \\
  F^{quant}(\mbf{q}) & \equiv  F(\mbf{q}) -  F^{clas}(\mbf{q}) \nonumber \\
  & = \left| f_{\textbf{0},\textbf{0}} (\mbf{q}) \right|^2 \sum\limits_{\mbf{n},\mbf{m}} 
  e^{ \imath \mbf{q} \left( \mbf{r}_{\mbf{n}} - \mbf{r}_{\mbf{m}} \right) } 
  \left( \aver{n_{\mbf{n}} n_{\mbf{m}}}- n^2 \right).
  \label{eqn:FQuantDef}
\end{align}
The former yields the classical amplitude of the scattered light 
$\left| \aver{a_{\mbf{k} \lambda}} \right|^{2}$, whereas the latter 
represents the remaining quantum contribution that together with $F^{clas}(\mbf{q})$ 
sum up to the total number of photons $\aver{a_{\mbf{k} \lambda}^\dagger a_{\mbf{k} \lambda}}$. 
We note that $F^{clas}(\mbf{q})$ has a form characteristic for a Bragg scattering 
and it is not affected by any statistical properties of the ultracold gas of bosons. 
On the contrary, $F^{quant}(\mbf{q})$ is sensitive to the atom number statistics and 
thus enables us to investigate statistical properties of different quantum states.

\section{Scattering from a Bose gas in one-dimensional optical lattice at zero temperature}
\label{sec:oneDimTzeroScatt}

\begin{figure}[b]
  \centering
  \includegraphics[scale=0.803]{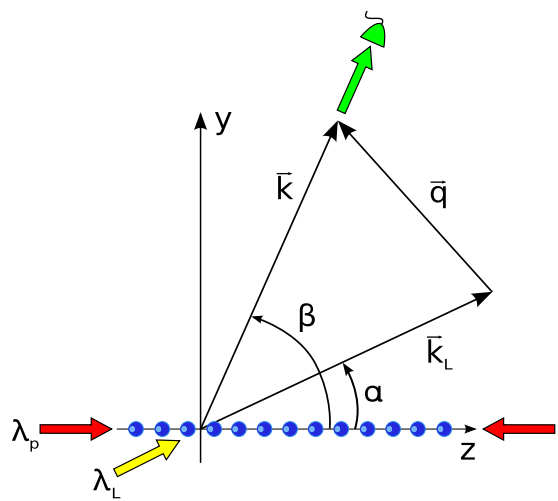}
  \caption{(Color online) Setup. A quasi one-dimensional optical lattice is 
    illuminated with a probing laser set at an angle $\alpha\,\epsilon\,[-\pi,\pi[$. 
    A detector is set at an angle $\beta\,\epsilon\,[-\pi,\pi[$.}
  \label{fig:System}
\end{figure}

The geometry of the system we investigate in this section is depicted in 
Fig. \ref{fig:System}. We consider one-dimensional homogeneous optical 
lattice generated by two overlapping and counterpropagating laser beams 
characterized by the wavelength $\lambda_p$. Atoms confined to the periodic 
potential are illuminated with a single laser beam with the wavelength $\lambda_{L}$. 
For the single particle basis that we have chosen the states
\begin{equation}
  \psi_{m}(\mbf{r}) = w_{m}(z) \,\psi_{\scriptscriptstyle{\perp}}(x,y)
\end{equation}
are products of a Wannier function $w_{m}(z)$ localized at lattice site $m$ 
along $z$-direction, and a Gaussian function $\psi_{\scriptscriptstyle{\perp}}(x,y)$ 
in tightly confined, perpendicular direction. For simplicity we assume the cylindrical 
symmetry  $\psi_{\scriptscriptstyle{\perp}}(x,y) = \psi_{\scriptscriptstyle{\perp}}(\rho)$.
At zero temparature a gas of bosons in a periodic potential appears in two distinct 
quantum phases \cite{fisher:1989,jaksch:1998}. When the tunneling process dominates 
over the on-site atom repulsion the system is found in the superfluid (SF) phase that 
is characterized by the presence of a global coherence and a non-zero order parameter. 
In contrast, for the on-site interactions stronger than the tunneling rate, the system 
exhibits the Mott-insulator phase. In the latter case the global coherence is lost, 
while the on-site particle number is fixed and the on-site fluctuations are suppressed.

For a Bose gas at zero temperature and deep in the MI regime, the on-site fluctuations 
and correlations vanish: $\aver{n_m n_{m'}} - \aver{n_m}\aver{n_{m'}}=0$. Hence, the 
quantum part $F^{quant}(\mbf{q})$ is zero identically, and the scattering is described 
by the standard Bragg pattern with characteristic set of maxima and minima, corresponding 
to the directions of constructive and destructive interference. In contrast, SF phase at 
$T=0$ exhibits nonzero fluctuations and correlations: 
$\aver{n_m n_{m'}} - \aver{n_m}\aver{n_{m'}} = n \delta_{m m'}- \frac{n^2}{N}$. Hence, 
apart from the similar behavior of the classical part $F^{clas}(\mbf{q})$ as for MI phase, 
the SF phase also gives rise to nonzero quantum component $F^{quant}(\mbf{q})$ which, within 
the deep lattice approximation (Eq.~\eqref{eqn:FdeepLatt}), is given by 
$F^{quant}(\mbf{q}) = N \left| f_{0,0} (\mbf{q}) \right|^2 \label{eqn:FSF0rd}$. 
This offers a unique possibility of a non-destructive measurement that allows one 
to distinguish between SF and MI phases \cite{Mekhov:2007}.

\begin{figure}[b]
  \centering
  \subfigure[\:$\alpha=\pi/4$]{
    \includegraphics[scale=0.39]{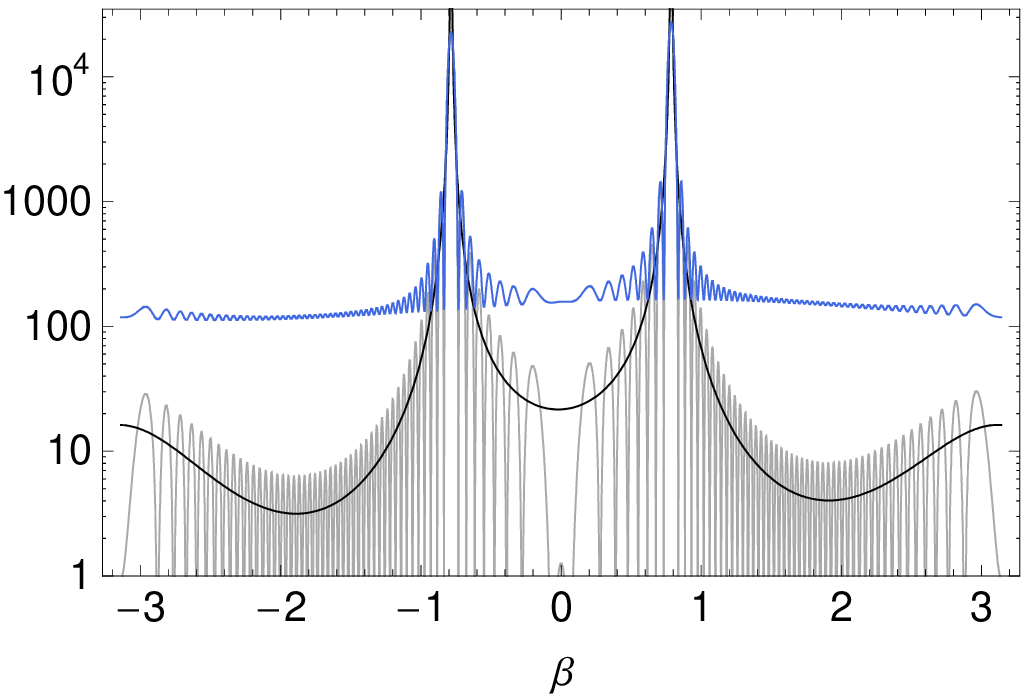}}
  \subfigure[\:$\alpha=\pi/2$]{
    \includegraphics[scale=0.39]{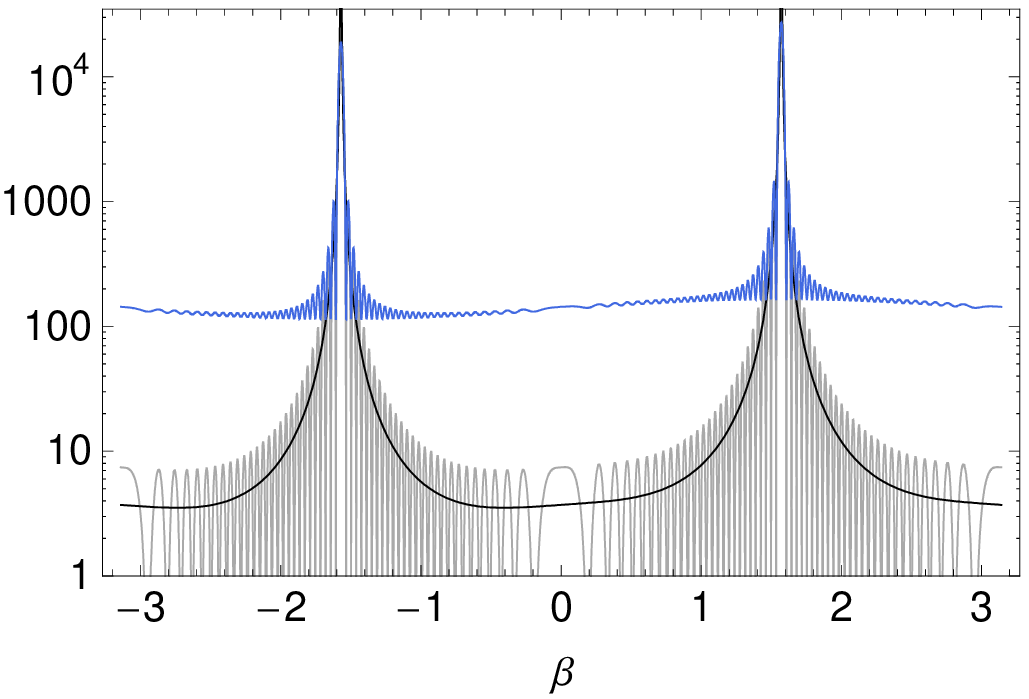}}
  \caption{(Color online) Structure function $F(\mbf{q})$ for SF (blue top curve) and 
    MI (gray bottom curve) phases, in the deep lattice approximation, for a probing 
    laser set at different angles $\alpha$, and a detector set at angle $\beta$. Here, 
    $V_0 = 15 E_r$, $M=55$, $N=3 M$, $\lambda_{p}/ \lambda_{L}=1$. The black line 
    represents average distribution (\ref{eqn:FMI0rdAvg}).}
  \label{fig:DiffA}
  \centering
  \subfigure[\:$\lambda_{p}/ \lambda_{L}=2$]{
    \includegraphics[scale=0.39]{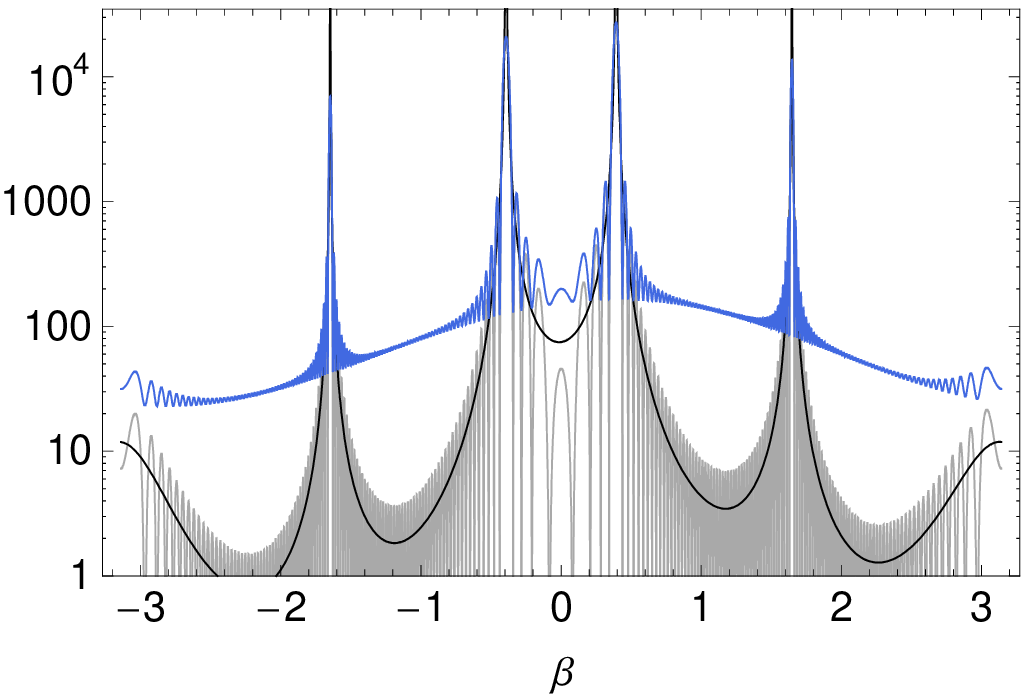}}
  \subfigure[\:$\lambda_{p}/ \lambda_{L}=1/3$]{
    \includegraphics[scale=0.39]{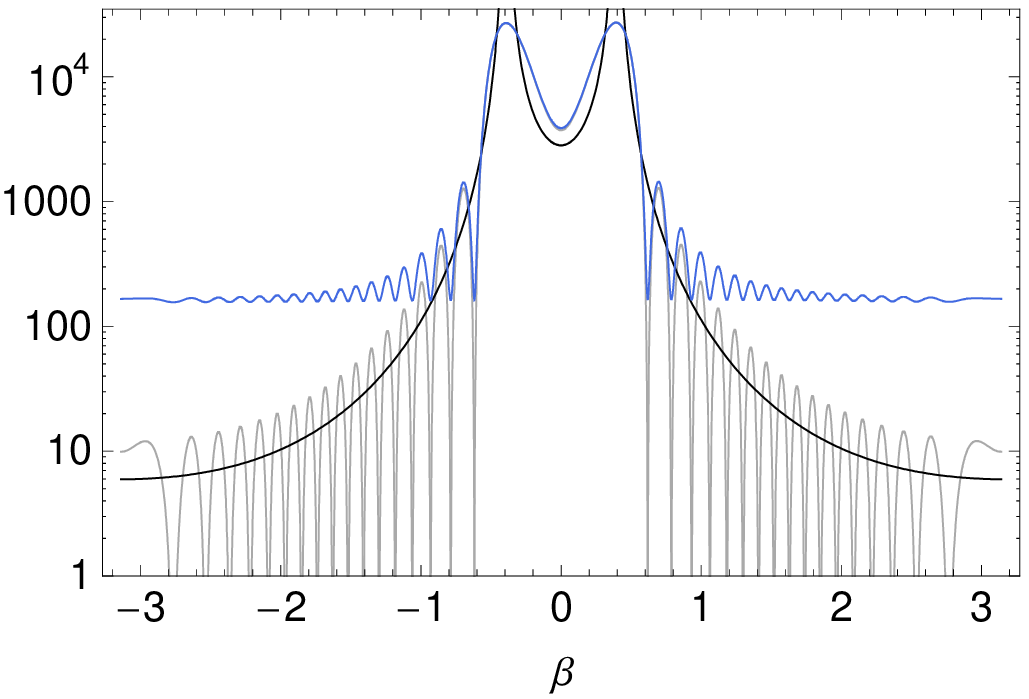}}
  \caption{(Color online)  Structure function $F(\mbf{q})$ for SF (blue top curve) 
    and MI (gray bottom curve) phases, in the deep lattice approximation, for 
    different ratios of $\lambda_{p} /\lambda_{L}$, and a detector set at angle $\beta$. 
    Here, $V_0 = 10 E_r$, $M=55$, $N=3 M$, $\alpha=\pi/8$. The black line represents the 
    average distribution (\ref{eqn:FMI0rdAvg}).}
  \label{fig:DiffL}
\end{figure}

Fig.~\ref{fig:DiffA} and Fig.~\ref{fig:DiffL} compare the scattering patterns from 
the SF and MI phases for the systems of $M=55$ sites with different configurations 
of the probing laser and different ratios of $\lambda_p$ to $\lambda_{L}$. At some 
characteristic angles corresponding to the Bragg scattering minima due to the 
destructive interference, the scattering from the MI state vanishes. In contrast, 
the scattering pattern from the SF state is nonzero at all angles, also in the 
directions where the classical component vanishes. We observe that a change of a 
ratio $\lambda_{p}/\lambda_{L}$ affects the scattering pattern, in particular a 
number and positions of the highest peaks resulting from the constructive interference.

We note that for large $M$ the scattering pattern quickly oscillates and thus, 
in the realistic measurement, one would detect photons scattered in some finite 
solid angle $d \Omega$ which is characteristic for the detector and that 
contains several interference fringes. Hence, we find it more appropriate 
to calculate the angular distribution of photons that is averaged over few 
neighboring maxima. The averaging does not affect the scattering pattern of 
SF phase, which is rather smooth, but it is important for MI phase. In 1D 
optical lattice, the angular distribution of photons scattered from MI state 
is determined by
\begin{align}
  F^{MI}(\mbf{q}) = \left| f_{0,0} (\mbf{q}) \right|^2 n^2 \frac{\sin^2 
    \left(\frac M 2 \mbf{q}\mbf{d} \right)}{\sin^2\left(\frac12 \mbf{q}\mbf{d} \right)}
  \label{eqn:FMI0rd}
\end{align}
where $\mbf{d}$ denotes the translation vector of a 1D lattice. Averaging over 
some finite solid angle around $\mbf{q}$ containing several maxima yields
\begin{align}
  \overline{F^{MI}(\mbf{q})} = \left| f_{0,0} (\mbf{q}) \right|^2 
  \frac{n^2}{2} \frac{1}{\sin^2\left(\frac12 \mbf{q}\mbf{d} \right)}.
  \label{eqn:FMI0rdAvg}
\end{align}
The above result is derived provided that the measurement is done not too close 
to the main maxima determined by the directions of the constructive interference. 
As can be observed in Fig.~\ref{fig:DiffA} and Fig.~\ref{fig:DiffL}, the 
averaged distribution of the light scattered from MI phase still can be well 
distinguished from the scattering from the SF phase, and result \eqref{eqn:FMI0rdAvg} 
for the averaged distribution remains approximately valid even close to the points of 
the destructive interference.

\begin{figure}[t]
  \centering
  \subfigure[\hspace{0.07cm} SF scattering at $V_0=1E_r$]{{
      \label{fig:poprawkiSF}}{{\includegraphics[scale=0.39]{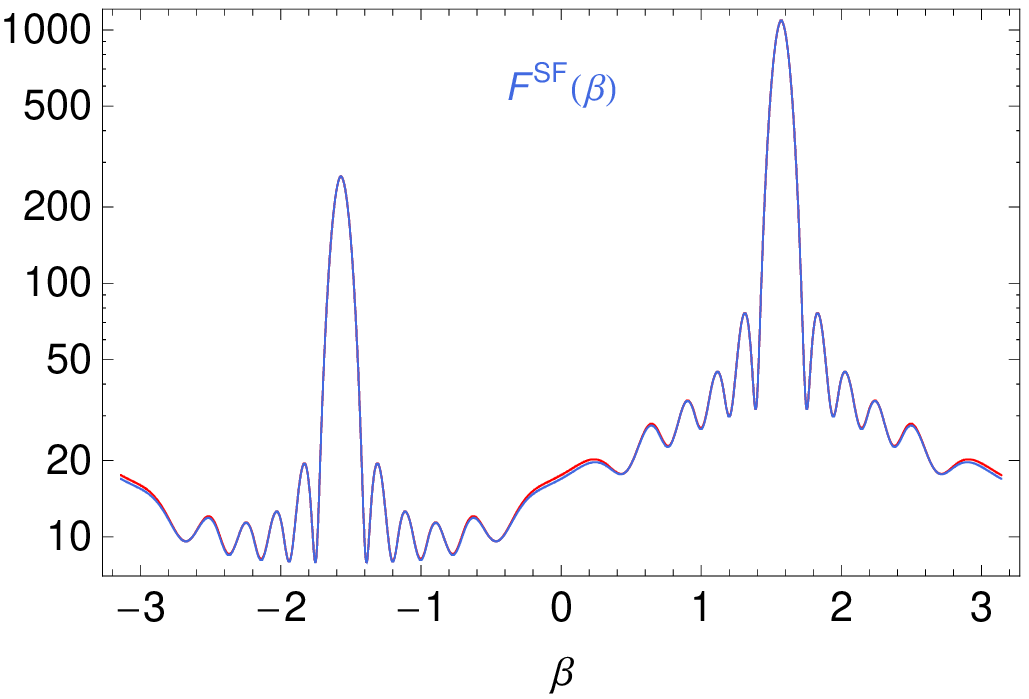}}}}
  \subfigure[\hspace{0.07cm} MI scattering at $V_0=5E_r$]{{
      \label{fig:poprawkiMI}}{{\includegraphics[scale=0.39]{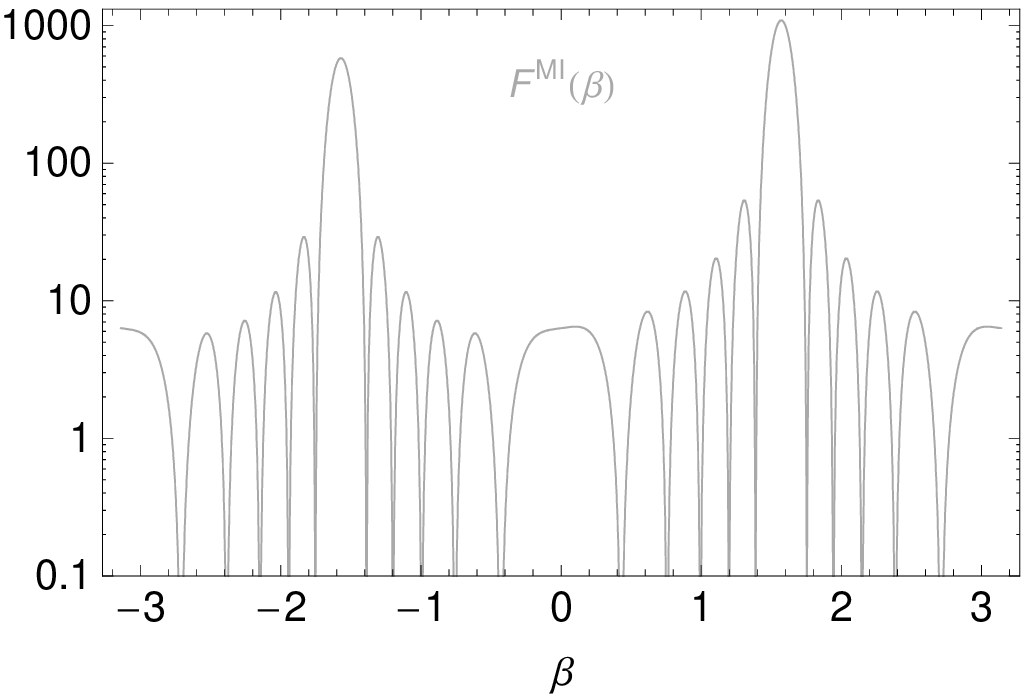}}}}
  \caption{
    (Color online) Distribution of light scattered from SF (left panel) and MI 
    (right panel) phases at zero temperature, versus the angle $\beta$. Here, 
    $M=11$, $N=3 M$, $\alpha = \pi/2$, and $\lambda_{p}/ \lambda_{L}=1$. The 
    red curves present $F(\beta)$ calculated in local approximation 
    \eqref{eqn:FdeepLatt}, whereas the blue curve (in the SF case) and the gray 
    curve (in the MI case) show results including also non-local Franck-Condon 
    factors as in Eq.~\eqref{eqn:Fq}. In the case of scattering from the SF 
    state at $V_0=1 E_r$ one notices slight differences between the two curves 
    in a vicinity of $\beta=0$ and $\beta= \pm \pi$. In the case of scattering 
    from the MI state at $V_0=5 E_{r}$ the two curves are indistinguishable.}
  \label{fig:poprawki}
\end{figure}

While performing the deep lattice approximation in Eq.~(\ref{eqn:FdeepLatt}) we 
have dropped all the Franck-Condon factors corresponding to transitions between states 
localized in different lattice sites. Obviously, with the decreasing lattice depth, the 
Wannier states begin to overlap between neighboring sites and we expect the nonlocal 
Franck-Condon factors to give larger contribution. In order to investigate this issue 
in Fig.~\ref{fig:poprawki} we compare the deep lattice approximation \eqref{eqn:FdeepLatt} 
with the result that includes summation over all pairs of the lattice sites in 
Eq.~\eqref{eqn:Fq}. For the clarity of presentation we show the results for a relatively 
small system of $M=11$ sites and a lattice depth $V=1 E_r$ and $V=5 E_r$ for SF and MI 
phases, respectively, expressed in the units of the recoil energy $E_r = \hbar^2 k_L^2/(2 m)$. 
We observe that even for the shallow lattice potential the nonlocal corrections give 
negligible contribution for the scattering from the SF state. Moreover, for the MI state 
the nonlocal corrections are even smaller because of the deeper lattice required to achieve 
this phase. In Appendix~\ref{app:nonlocal} we show that corrections due to the nearest 
neighbors in weak lattices are isotropic and scale as the total number of atoms $N$.
Therefore, nonlocal corrections give rise to a scattering at the angles of destructive 
interference of the classical part. However, for typical lattice depths, the corresponding 
contribution is small and can be totally neglected for both quantum phases.

\section{Statistical properties at finite temperatures}
\label{sec:statProperties}

\subsection{Bose-Hubbard model of an ultracold gas in a periodic potential}

As discussed in the previous section, the angular distribution of the 
scattered light is determined by the occupation number statistics in 
lattice sites. We investigate the occupation number statistics within 
Bose-Hubbard (BH) model \cite{fisher:1989,jaksch:1998}, considering only 
excitations within the lowest Bloch band. The BH Hamiltonian reads:
\begin{equation}
  \Ham = -J \!\!\!\!\sum\limits_{\aver{\mbf{m},\mbf{m}'}} g^{\dagger}_{\mbf{m}} g_{\mbf{m}'}
  +\frac{1}{2}U\!\sum\limits_{\mbf{m}} \nop_{\mbf{m}} \left( \nop_{\mbf{m}}-1 \right)
  \label{eqn:BHHam}
\end{equation}
where the first sum on the right-hand side is restricted to nearest 
neighbors only. The parameter
\begin{equation}
  J \equiv -\int d^{3}r \:w^{\ast}_{\mbf{m}} (\mbf{r})\left[  -\frac{\hbar^2}{2m}\nabla^{2}+
    V_{p} (\mbf{r}) \right] w_{\mbf{m}'}(\mbf{r})
\end{equation}
is the hopping matrix element between neighboring sites $\mbf{m}$ 
and $\mbf{m}'$, and parameter:
\begin{equation}
  U \equiv \frac{4 \pi a_s \hbar^2}{m} \int d^{3}r \left| w_{\mbf{m}}(\mbf{r}) \right|^4
\end{equation}
corresponds to the strength of the on site repulsion of two atoms in 
a lattice site $\mbf{m}$. As before, $w_{\mbf{m}}(\mbf{r})$ is the 
single-particle Wannier's wavefunction of an atom occupying site $\mbf{m}$ 
in a lattice.

The BH Hamiltonian can be equivalently expressed in the momentum space, 
which is convenient for the analysis of the SF phase at finite 
temperatures and application of the Bogoliubov method. To this end 
we introduce annihilation and creation operators in momentum space
\begin{align}
  a_{\mbf{k}} = &\frac{1}{\sqrt{M}} \sum\limits_{\mbf{m}} g_{\mbf{m}} 
  e^{\imath \mbf{k} \mbf{r}_{\mbf{m}}} \label{eqn:anniFour}, \\
  a^{\dagger}_{\mbf{k}} = &\frac{1}{\sqrt{M}} \sum\limits_{\mbf{m}} 
  g^{\dagger}_{\mbf{m}} e^{-\imath \mbf{k} \mbf{r}_{\mbf{m}}}, 
  \label{eqn:creatFour}
\end{align}
respectively, in which index $\mbf{m}$ runs over all sites in a lattice. 
A period of the cubic lattice is $d=\lambda_{p}/2$ and a size of the 
system is equal to $L=M^{1/3} d$. The periodic boundary conditions imply 
quantization of a wave vector: $\; \mbf{k} = \frac{2 \pi}{L} (n_x,n_y,n_z)$, 
where $n_i$ are integer numbers ranging from $-\lfloor M/2 \rfloor$ to 
$\lfloor M/2 \rfloor$ \cite{rem2}. By rewriting Eq.~(\ref{eqn:BHHam}) in terms of 
$a_{\mbf{k}}$ and $a^{\dagger}_{\mbf{k}}$, we find:
\begin{align}
  \Ham =& \sum\limits_{\mbf{k}} \epsilon_{\mbf{k}} a^{\dagger}_{\mbf{k}} a_{\mbf{k}}
  + \frac{U}{2 M} \sum\limits_{\mbf{k}, \mbf{k}', \mbf{k}''} a^{\dagger}_{\mbf{k} + 
    \mbf{k}''} a^{\dagger}_{\mbf{k} '- \mbf{k}''} a_{\mbf{k} '} a_{\mbf{k}}
  \label{eqn:BHHamM}
\end{align}
where
\begin{equation}
  \epsilon_{\mbf{k}} \equiv 6 J - 2 J \sum\limits_{i=1}^{3} \cos\left( k_i d \right).
\end{equation}

\subsection{Statistical properties of the superfluid phase}

The standard description of a weakly interacting Bose gas is 
based on the Bogoliubov approximation \cite{bogoliubov:1947} 
and can be also applied for a superfluid phase in periodic 
potentials \cite{stoof:2001}. We perform the Bogoliubov 
approximation to Hamiltonian (\ref{eqn:BHHamM}), replacing 
the annihilation and creation operators in the zero 
quasi-momentum modes ($\mbf{k}=0$) by $\mathbb{C}$-numbers 
$a_0 \approx a_0^\dagger \approx \sqrt{N_0}$. By introducing 
the quasi-particle annihilation and creation operators 
$b_{\mbf{k}}$ and $b^{\dagger}_{\mbf{k}}$, respectively, 
which fulfill the standard bosonic commutation rules 
$\left[b_{\mbf{k}}, b^{\dagger}_{\mbf{k}'}\right] = \delta_{\mbf{k},\mbf{k}'}$, 
and are related to $a_{\mbf{k}}$ and $a^{\dagger}_{\mbf{k}}$ 
by the canonical transformation
\begin{equation}
  \begin{pmatrix} b_{\mbf{k}}\\b^{\dagger}_{-\mbf{k}} \end{pmatrix} =
  \begin{pmatrix} u_{\mbf{k}} & v_{\mbf{k}}\\v_{\mbf{k}} & u_{\mbf{k}}\end{pmatrix}
  \begin{pmatrix} a_{\mbf{k}}\\a^{\dagger}_{-\mbf{k}} \end{pmatrix},
  \label{eqn:botranf}
\end{equation}
we diagonalize Hamiltonian (\ref{eqn:BHHamM}) obtaining
\begin{equation}
  \Ham = E_0 + \sum\limits_{\mbf{k}} \hbar \omega_{\mbf{k}} 
  b^{\dagger}_{\mbf{k}} b_{\mbf{k}}.
  \label{eqn:BogHamDiag}
\end{equation}
Here, $E_0$ represents constant, ground-state energy term, 
$\hbar \omega_{\mbf{k}}$ are the energies of the quasi-particle 
excitation spectrum
\begin{align}
  \hbar \omega_{\mbf{k}} = \sqrt{\epsilon_{\mbf{k}}^{2} + 
    2 U \frac{N_0}{M} \epsilon_{\mbf{k}}},
\end{align}
and real-valued coefficients $u_{\mbf{k}}$ and 
$v_{\mbf{k}}$ of the Bogoliubov transformation are given by
\begin{equation}
  v_{\mbf{k}}^{2} = u_{\mbf{k}}^{2} - 1 = \frac{1}{2} 
  \left( \frac{\epsilon_{\mbf{k}} + 
      U \frac{N_0}{M}}{\hbar \omega_{\mbf{k}}} -1 \right).
\end{equation}
We note, that the excitation spectrum $\hbar \omega_{\mbf{k}}$ 
depends on the condensate population $N_0$, which is known 
in the literature as the Bogoliubov-Popov spectrum, and 
is well suited to describe the statistics of a BEC at finite 
temperatures \cite{Svidzynsky,Idziaszek3}.

Below the critical temperature, when the condensate is 
macroscopically occupied, the occupation statistics of the 
quasi-particle modes is given by the Bose-Einstein distribution:
\begin{align}
  \label{eqn:fk}
  \aver{b^{\dagger}_{\mbf{k}} b_{\mbf{k}}} = & \frac{1}{e^{\beta \hbar 
      \omega_{\mbf{k}}}-1} \equiv f_\mbf{k},\quad\mbf{k}\neq 0,\\
  \label{eqn:fk2}
  \aver{b^{\dagger}_{\mbf{k}} b_{\mbf{k}} b^{\dagger}_{\mbf{k}'} b_{\mbf{k}'}} = & 
  f_\mbf{k} f_{\mbf{k}'} + \delta_{\mbf{k},\mbf{k}'} \left( f_\mbf{k}^2 + 
    f_{\mbf{k}} \right), \quad\mbf{k},\mbf{k}'\neq 0,
\end{align}
where the value of the chemical potential $\mu$ is set to zero. 
This follows from the fact the condensate acts as a reservoir 
of particles, and distributions of particles in excited modes 
are not restricted by the particle number conservation, which is 
consistent with the so-called Maxwell-Demon (MD) ensemble approximation
\cite{navez:1997,grossman:1997,wilkens:1997}. Applying 
Bogoliubov transformation \eqref{eqn:botranf}, and 
Eqs.~(\ref{eqn:fk}) and (\ref{eqn:fk2}) we can easily find 
the mean occupation, fluctuations and correlations of the 
number of atoms in the quantized quasi-momentum modes:
\begin{align}
  \aver{n_{\mbf{k}}} =& (u_\mbf{k}^2 + v_\mbf{k}^2) f_{\mbf{k}} + v_\mbf{k}^2, 
  \label{eqn:nk} \\
  \aver{\delta^{2} n_{\mbf{k}}} =& \left( u_{\mbf{k}}^{2} + v_{\mbf{k}}^{2} \right)^2
  \left( f_{\mbf{k}}^{2} + f_{\mbf{k}} \right) + u_{\mbf{k}}^2 v_{\mbf{k}}^2, \\
  \aver{n_{\mbf{k}} n_{\mbf{k}'}} = & u_{\mbf{k}}^{2} v_{\mbf{k}}^{2} 
  \left( 1 + 4 f_{\mbf{k}} + 4 f_{\mbf{k}}^{2} \right)
  \delta_{\mbf{k},-\mbf{k}'} + \aver{n_{\mbf{k}}} \!\aver{n_{\mbf{k}'}} 
  \label{eqn:nknkp}
\end{align}
in which $\mbf{k} \neq \mbf{k}' \neq \mbf{0}$ and 
$n_{\mbf{k}} \equiv a^{\dagger}_{\mbf{k}} a_{\mbf{k}}$ 
is a particle number operator for a quasi-momentum mode 
$\mbf{k}$. Calculations of statistical quantities 
\eqref{eqn:nk}-\eqref{eqn:nknkp} within the Bogoliubov-Popov 
method require self-consistent determination of the mean 
condensate population $N_0$. First, $N_0$ enters the 
excitation spectrum as a parameter. Second, it is determined 
by the statistics itself,
\begin{equation}
  N_0 = N - \sum_{\mbf{k} \neq \mbf{0}} \aver{n_{\mbf{k}}},
\end{equation}
which yields
\begin{align}
  N_0 = N - & \sum\limits_{\mbf{k} \neq 0} \left( \frac{\epsilon_{\mbf{k}}
      + U \frac{N_0}{M}}{\hbar \omega_{\mbf{k}}} f_{\mbf{k}} + 
    \frac{\epsilon_{\mbf{k}} + U \frac{N_0}{M} - \hbar \omega_{\mbf{k}}}
    {2 \hbar \omega_{\mbf{k}}} \right).
  \label{eqn:BOgN0}
\end{align}
\begin{figure}[t]
  \centering
  \subfigure[\:Number of condensate atoms in the lattice.]{{\label{fig:SFcomboStatisticsA}}{{
        \includegraphics[scale=0.39]{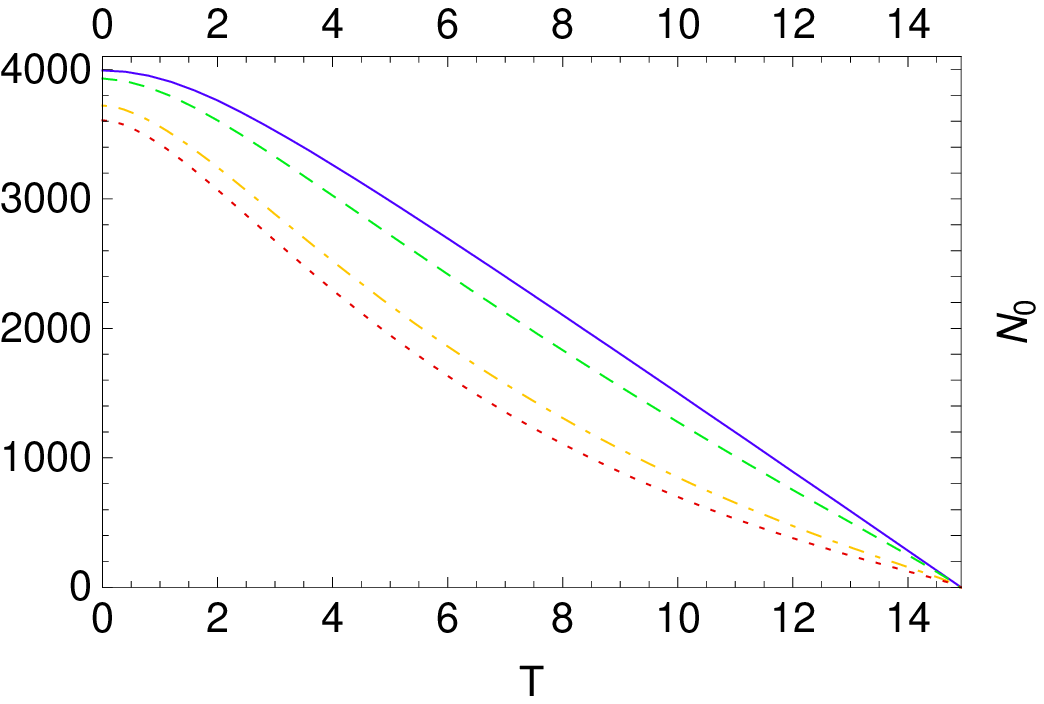}}}}
  \subfigure[\:Atoms number fluctuations in a single site of the lattice.]{{\label{fig:SFcomboStatisticsB}}{{
        \includegraphics[scale=0.38]{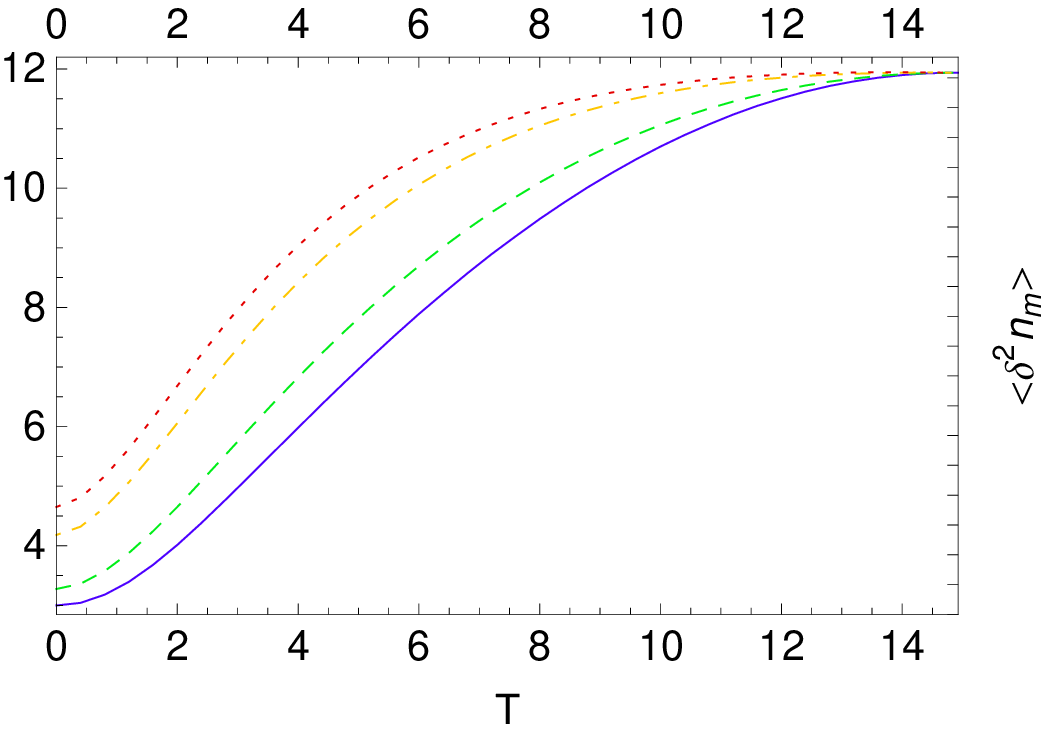}}}}\\
  \subfigure[\:Correlations between close neighbours. Here, $\mbf{m} = (0,0,0)$ and
  $\mbf{m}' =(1,0,0)$.]{{\label{fig:SFcomboStatisticsC}}{{
        \includegraphics[scale=0.38]{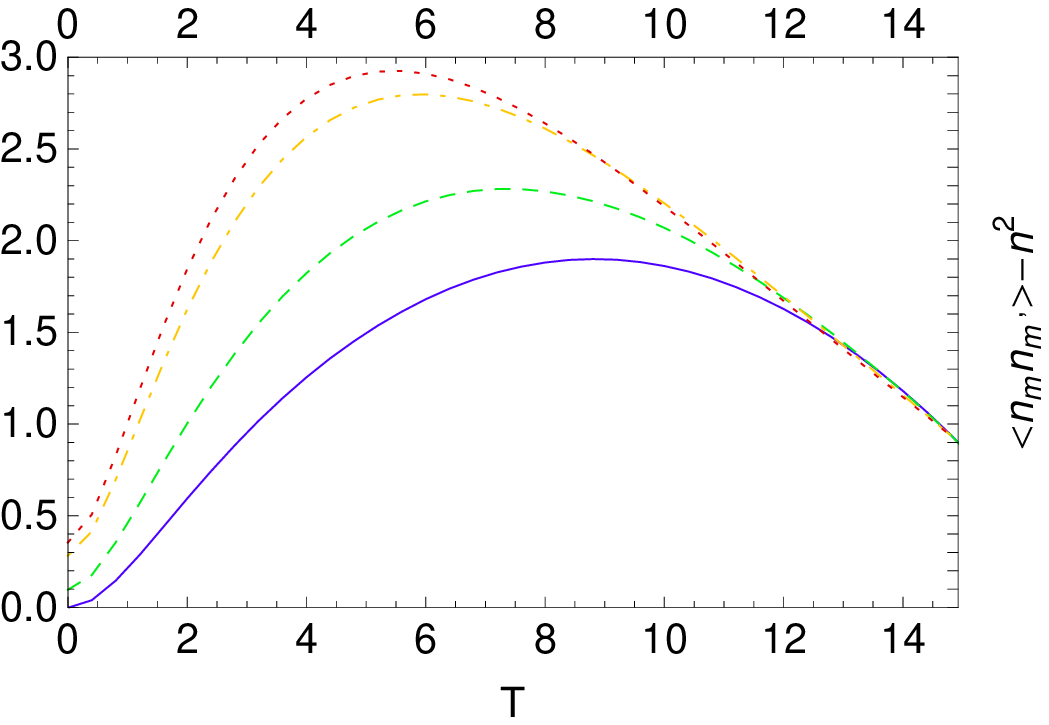}}}}
  \subfigure[\:Correlations between distant neighbours. Here, $\mbf{m} = (0,0,0)$ and
  $\mbf{m}' =(5,5,5)$.]{{\label{fig:SFcomboStatisticsD}}{{
        \includegraphics[scale=0.39]{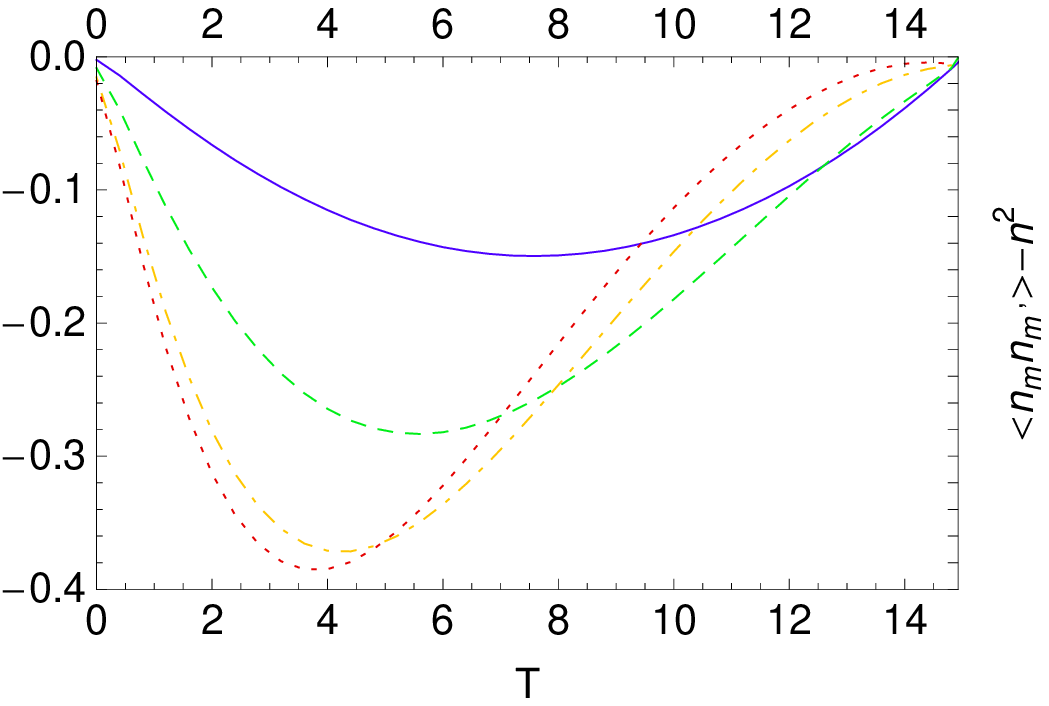}}}} \\
  \subfigure[\:Momemtum modes correlations. Here, $\mbf{k}=(2,2,2).$]{{\label{fig:SFmomSpaceStatisticsE}}{{
        \includegraphics[scale=0.39]{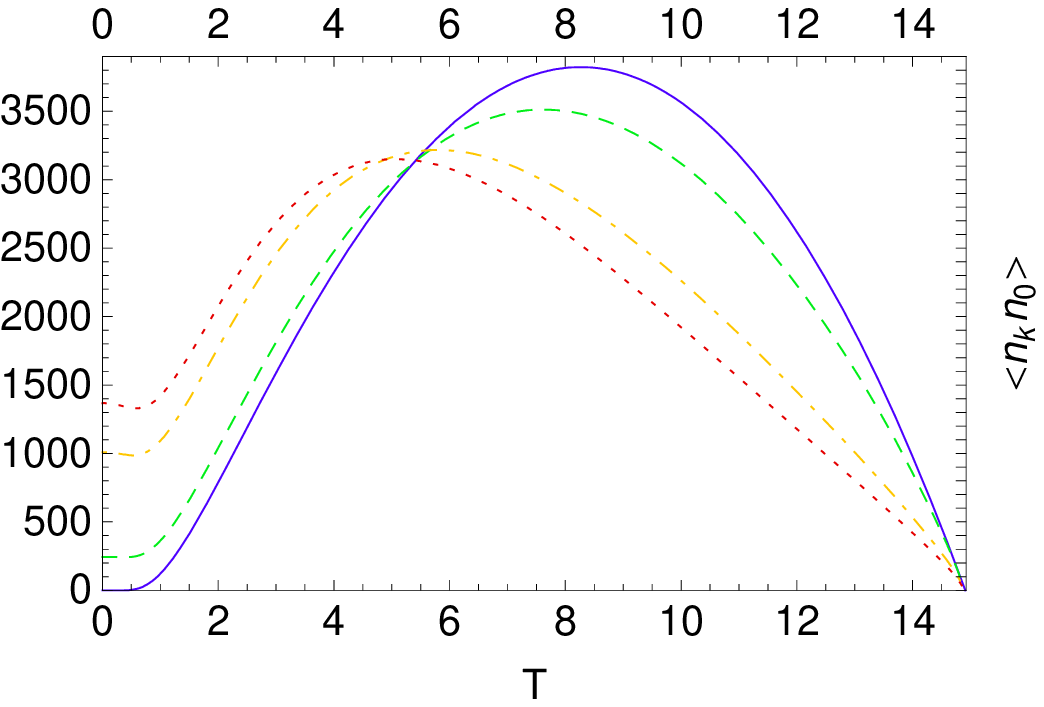}}}}
  \subfigure[\:Momemtum modes correlations. Here, $\mbf{k}=(1,0,0)$ and 
  $\mbf{k}' = (1,1,1).$]{{\label{fig:SFmomSpaceStatisticsF}}{{
        \includegraphics[scale=0.39]{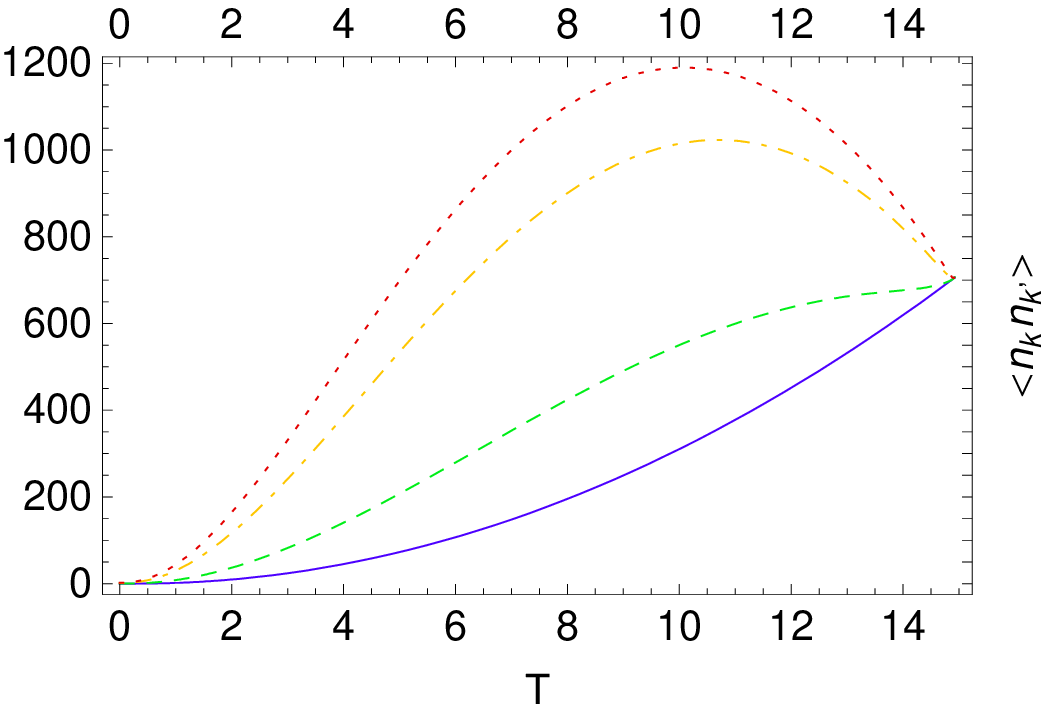}}}}
  \caption{(Color online) Statistical properties of the SF phase in an optical
    lattice. Here, $M=11 \!\times\! 11 \!\times\! 11$ and $N= 3 M$. 
    Different colors of the curves refer to different values of the 
    parameter $U$: blue (solid) for $U=0$, green (dashed) for $U=1$, 
    yellow (dot-dashed) for $U=4$, red (dotted) for $U=6$. The selected 
    values of $U$ imply the following values of quantum depletion: 
    0\%, 1.6\%, 6.8\%, 9.7\%, respectively. The strength of interaction 
    $U$ and temperature $T$ are expressed in units of~$J$.}
  \label{fig:SFcomboStatistics}
\end{figure}
Finally, by transforming to the position space with the help of 
Eqs.~(\ref{eqn:anniFour}) and (\ref{eqn:creatFour}), we evaluate 
single-site occupation number statistics, i.e. single-site 
fluctuations, $\aver{\delta^2 n_{\mbf{m}}} = \aver{n_{\mbf{m}}^2} - n^2$, 
and correlations, $\aver{n_{\mbf{m}} n_{\mbf{m}'}} - n^2$, between each pair 
of sites in the lattice. At the final stage they are substituted 
into Eq.~(\ref{eqn:FdeepLatt}) determining the angular 
distribution of the scattered light.

In Fig. \ref{fig:SFcomboStatistics} we present results 
for the statistics of the SF state realized in an optical 
lattice. We have chosen four different values of the strength of 
interactions $U$ that correspond to quantum depletion 
ranging from $0$ to approximately $0.1$, which should be proper in 
the regime of a weakly interacting gas where the Bogoliubov 
method is applicable. An upper limit of temperatures we 
consider is established by the conditions of validity of the MD 
ensemble approximation that, for sufficiently large systems, works 
well up to a temperature close to the critical temperature $T_C$. 
As expected, we observe that the on-site fluctuations 
increase monotonically up to $T_C$. In contrast, the correlations 
between populations of different sites exhibit non-monotonic behavior 
that is strongly dependent on a distance between the considered sites. 
The phenomena can be understood by studying the behavior of 
these statistical quantities at small and at large temperatures. 
Readily, in the limit $T \rightarrow 0$ the on-site fluctuations and 
correlations follow the behavior presented in Section 
\ref{sec:oneDimTzeroScatt}: 
$\aver{\delta^2 n_{\mbf{m}}} = n - \frac{n^2}{N}$ and 
$\aver{n_{\mbf{m}} n_{\mbf{m}'}} - n^2 = -\frac{n^2}{N}$. 
On the contrary, at large temperatures the fluctuations 
and correlations can be described consistently within a model 
of $N$ indistinguishable particles distributed over $M$ 
degenerate levels: $\aver{\delta^2 n_{\mbf{m}}} = n^2 + n$ 
and $\aver{n_{\mbf{m}} n_{\mbf{m}'}} - n^2 =0$. 
For both of the limits, the analytical expressions are derived 
under the assumption $U/J \rightarrow 0$, however the 
approximations work reasonably well also for finite values 
of the ratio $U/J$. The last two panels of Fig. \ref{fig:SFcomboStatistics} 
present correlations between different modes in the momentum space. 
We note that the correlation between an excited and the condensate 
mode $\aver{n_{\mbf{k}} n_{\mbf{0}}}$, exhibits a maximum at some 
moderate temperature. This follows simply from the competition 
between the process of thermal depletion of the condensate 
and a growth of the thermal fraction. Similarly, in case of 
correlations between two excited modes, we observe that at 
some temperature the initial growth of $\aver{n_{\mbf{k}} n_{\mbf{k}'}}$ 
is suppressed by decrease in population of these modes in 
favor of population of modes of some higher quasi-momentum.

\subsection{Statistical properties of the Mott-insulator phase}

We introduce grand canonical Bose-Hubbard Hamiltonian ${\cal K}$:
\begin{equation}
  {\cal K} = -J \!\!\!\!\sum\limits_{\aver{\mbf{m},\mbf{m}'}} 
  g^{\dagger}_{\mbf{m}} g_{\mbf{m}'}
  +\frac{1}{2}U\!\sum\limits_{\mbf{m}} \nop_{\mbf{m}} 
  \left( \nop_{\mbf{m}}-1 \right) - \mu \sum_{\mbf{m}} \nop_{\mbf{m}}.
  \label{eqn:GCbhHam}
\end{equation}
In order to describe quantum statistics of the Mott insulator phase at 
finite temperatures we adopt a mean-field decoupling approximation 
\cite{sachdev:1999,stoof:2003}. In analogy to the Bogoliubov approach, 
we introduce a complex mean-field parameter $\psi \equiv \aver{g_m}$ 
that can be physically interpreted as an order parameter that is 
nonzero if the system is superfluid. Below the phase transition 
point, the symmetry related to the gauge invariance of the phase 
is spontaneously broken and without losing generality we can assume 
that $\psi$ is real. The new parameter allows one to decouple 
the hopping term occurring in Eq.~(\ref{eqn:GCbhHam})
\begin{equation}
  g^{\dagger}_{\mbf{m}} g_{\mbf{m}'} = \psi \left( g^{\dagger}_{\mbf{m}}
    +  g_{\mbf{m}'} \right) - \psi^{2}.
  \label{eqn:midec}
\end{equation}
By performing this substitution we can decompose Hamiltonian~(\ref{eqn:GCbhHam}) 
into a sum of mean-field local Hamiltonians ${\cal K}^{\scriptscriptstyle{MF}}_{\mbf{m}}$:
$\displaystyle {\cal K} = \sum_{\mbf{m}} {\cal K}^{\scriptscriptstyle{MF}}_{\mbf{m}}$,
where
\begin{align}
  {\cal K}^{\scriptscriptstyle{MF}}_{\mbf{m}} \equiv & -2 D J \psi \left( g_{\mbf{m}}
    + g^{\dagger}_{\mbf{m}} \right) + 2 D J\psi^2 - \mu \nop_{\mbf{m}} \nonumber \\
  & +\frac{1}{2}U \nop_{\mbf{m}} \left( \nop_{\mbf{m}}-1 \right).
  \label{eqn:hamMF}
\end{align}
At zero temperature, calculation of the ground-state energy and its 
minimization as a function of the superfluid order parameter $\psi$ 
yields the phase diagram analytically \cite{stoof:2001}. However, 
for non-zero temperatures the model has no analytical solution, 
and one has to resort to numerical calculations. Namely, by 
diagonalization of Eq.~(\ref{eqn:hamMF}) we calculate grand 
canonical partition function $\mathcal{Z}(\psi)$,
\begin{equation}
  \mathcal{Z}(\psi) = \textrm{Tr} \left\{ e^{-\beta 
      {\cal K}^{\scriptscriptstyle{MF}}_{\mbf{m}}} \right\},
\end{equation}
and on its grounds we determine the grand thermodynamic 
potential $\Omega (\psi)$,
\begin{equation}
  \Omega (\psi)= -\frac{1}{\beta} \ln \mathcal{Z}(\psi).
\end{equation}
Subsequently, by minimizing $\Omega(\psi)$ with respect 
to $\psi$, we obtain the equilibrium value of the order parameter 
that we use to calculate all the relevant thermodynamic
quantities. In particular:
\begin{align}
  \aver{n_{\mbf{m}}(\mu)} =& \frac{1}{\mathcal{Z}} \textrm{Tr} 
  \left\{ \nop_{\mbf{m}} e^{-\beta {\cal K}^{\scriptscriptstyle{MF}}_{\mbf{m}}} \right\} 
  \label{eqn:nAverMI}, \\
  \aver{\delta^2 n_{\mbf{m}}(\mu)} =& \frac{1}{\mathcal{Z}} \textrm{Tr}
  \left\{ \nop^{2}_{\mbf{m}} e^{-\beta {\cal K}^{\scriptscriptstyle{MF}}_{\mbf{m}}} \right\} 
  - n^{2}.
\end{align}
In the homogeneous lattice $\aver{n_{\mbf{m}}(\mu)} = n$, and this 
identity is used to determine the value of the chemical potential, 
for a given single site occupation $n$.

Our mean-field approach assumes decoupling of 
different sites, $\aver{n_{\mbf{m}} n_{\mbf{m}'}} - n^2 =0$ ($\mbf{m} 
\neq \mbf{m}'$), that agrees with the zero temperature statistics of the 
MI assumed in section IV. This, in general, is not valid at higher 
temperatures when the hopping between adjacent sites 
in non-negligible. Nevertheless, it is satisfied at 
smaller temperatures considered here.

\begin{figure}[t]
  \centering
  \subfigure[\:Average atoms number in a single site in the lattice.]{
    \includegraphics[scale=0.376]{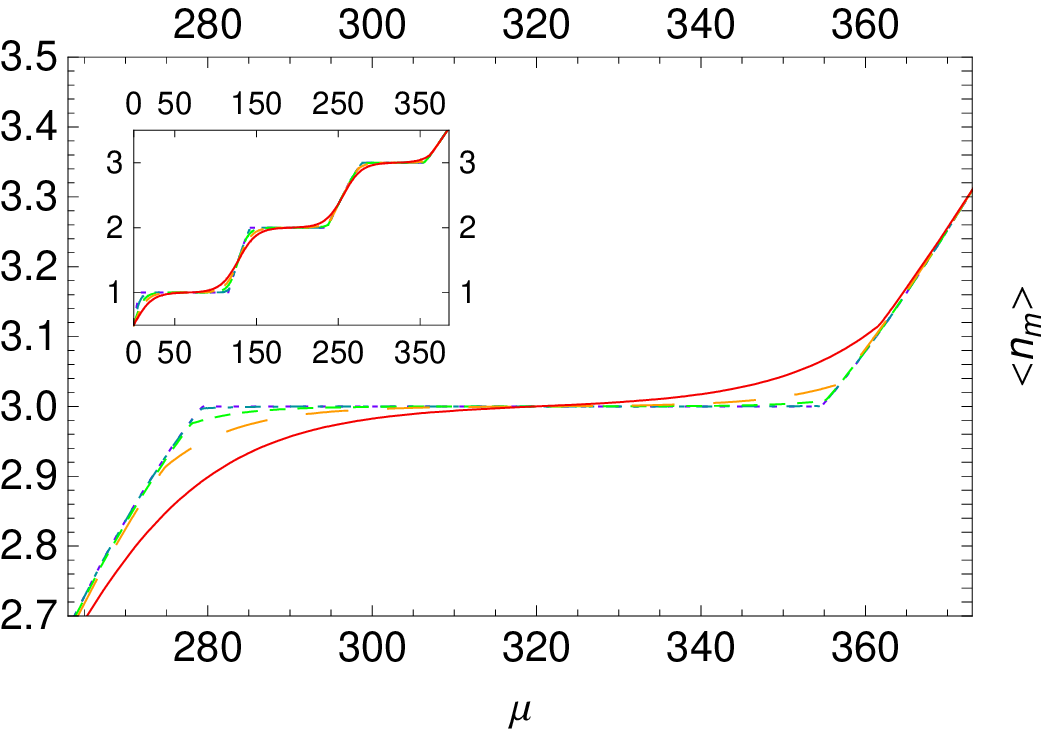}}
  \subfigure[\:Atoms number fluctuations in a single site in the lattice.]{
    \includegraphics[scale=0.39]{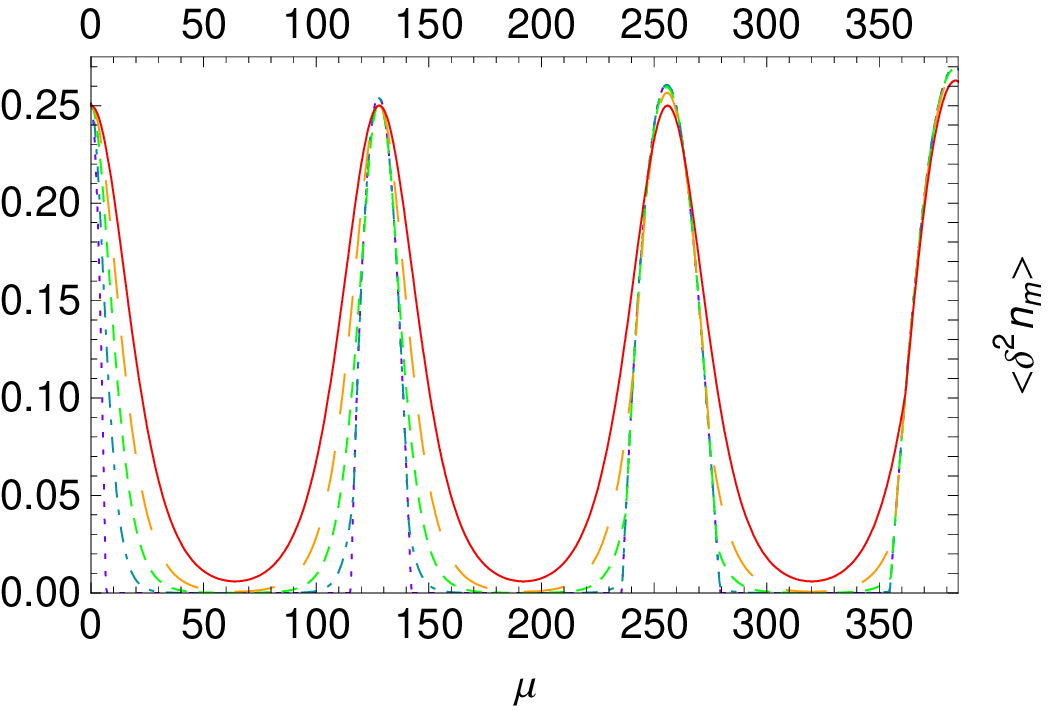}}
  \caption{(Color online) Statistical properties of Mott insulator phase 
    for $U=128$. Colors of the curves refer to different values of 
    temperature being considered: blue (dotted) for $T=0$,
    dark green (dot-dashed) for $T=4$, green (dashed) for $T=6$, 
    orange (long-dashed) for $T=8$ and red (solid) for $T=11$.
    All the values of parameters are expressed in units of $J$.}
  \label{fig:statsMI}
\end{figure}

In Fig. \ref{fig:statsMI} we present the mean single-site 
occupation and fluctuations as a function of the chemical potential, 
and for different system temperatures. The calculations have been performed 
within the mean-field model for $U/J = 128$. One can see the existence of a 
characteristic temperature above which, the flat steps in 
$\aver{n_{\mbf{m}}(\mu)}$ disappear completely, and the curve 
becomes monotonically increasing. This crossover is accompanied 
by an appearance of nonzero fluctuations for 
all values of $\mu$ presented in the plot. This corresponds to 
the transition from the MI to the normal phase. In MI phase the 
system is infinitely compressible: $\partial \aver{n}/\partial \mu = 0$, 
while in the normal or SF phase the compressibility becomes finite: 
$\partial \aver{n}/\partial \mu \neq 0$. Since the on-site fluctuations can be 
expressed as $\langle \delta^2 n \rangle = \partial \aver{n}/\partial(\beta\mu)$, 
therefore they can be nonzero only in the normal or SF phase. In order to 
distinguish between the normal and SF phases one can resort to the 
value of the order parameter $\psi$. Finally, we note that our 
mean-field treatment neglects the effects of 
correlations between different sites and the quantum fluctuations. In the more 
accurate models that take these effects into account, the on-site fluctuations 
become nonzero already in the MI regime, close to the boundaries with the SF or 
normal phases.

\section{Finite temperature scattering in three dimensions}
\label{sec:3D}

\begin{figure}[b]
  \centering
  \includegraphics[scale=0.41]{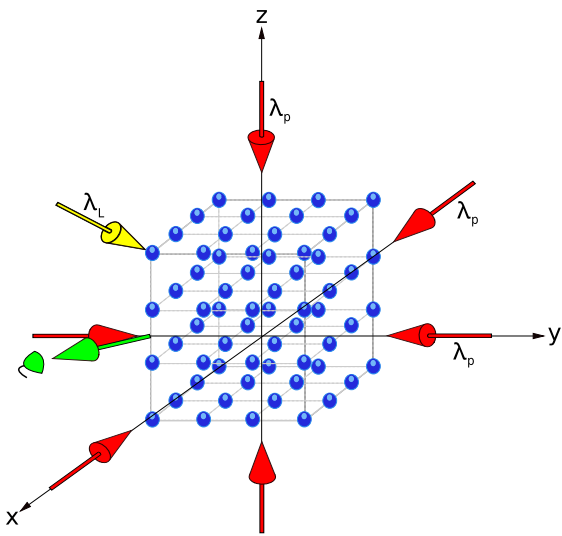}
  \caption{(Color online) Setup. A three-dimensional optical cubic lattice 
    generated by lasers $\lambda_p$ (red arrows) is illuminated by a probing 
    laser $\lambda_{L}$ (yellow arrow) set at angles $({\phi}_L, {\theta}_L)$. 
    A detector of scattered photons (green arrow) is aligned in a direction 
    $({\phi}_d, {\theta}_d)$.}
  \label{fig:3dsystem}
\end{figure}

We consider three-dimensional cubic lattice and assume a sufficiently 
large value of the trapping potential depth $V_0$ to neglect 
corrections from the nonlocal Franck-Condon factors. The setup 
is depicted in Fig.~\ref{fig:3dsystem}. The lasers creating an optical 
lattice ($\lambda_p$, red arrows) are set along $x,y$ and $z$ axes. 
In general, the position of a probing laser ($\lambda_{L}$, yellow arrow, 
characterized by angles $\phi_L$, $\theta_L$) and detector (green arrow, 
characterized by angles $\phi_d$, $\theta_d$) can be optimized in order to 
minimize the contribution from the classical component in the vicinity of the 
direction of the measurement, cf. Appendix \ref{app:optimization}. This is of 
particular importance in case of large lattices for which, due to a big number 
of interference fringes, the detector would collect the photons from several 
interference peaks. Here, though, we do not choose the optimal configuration, 
we consider some example geometry which not only sufficiently reduces an 
influence of the classical component but also offers relatively simple 
experimental realization. Namely, we choose the direction of the probing 
beam along one of the diagonals of the lattice cube, 
$\mbf{k}_L = |\mbf{k}_L| (\frac{1}{\sqrt{3}},-\frac{1}{\sqrt{3}},
\frac{1}{\sqrt{3}})$, and a detector centered around $\mbf{k} = 
|\mbf{k}_L| (\frac{1}{\sqrt{2}},-\frac{1}{\sqrt{2}},0)$.
\begin{figure}[b]
  \centering
  \includegraphics[scale=0.78]{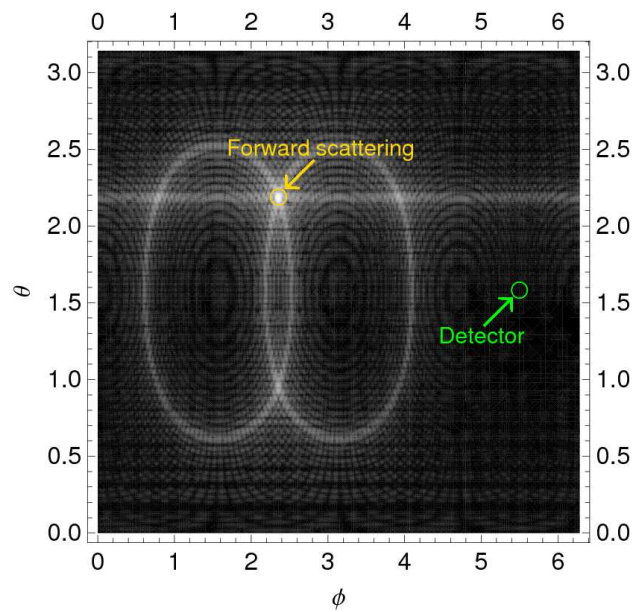}
  \caption{(Color online) Logarithm of the structure function $F(\mbf{q})$ 
    for MI phase at zero temperature, as a function of spherical angles of 
    detection, $(\theta,\phi)$. The probing laser is set at $({\phi}_L, {\theta}_L)$. Bright regions 
    correspond to directions in which a large number of photons is scattered. 
    The yellow circle (pointed by the yellow arrow) indicates the direction of 
    the probing laser (global maximum of number of scattered photons). The green 
    circle (pointed by the green arrow) refers to the direction of a detection 
    $\mbf{k} = |\mbf{k}_L| (\frac{1}{\sqrt{2}},-\frac{1}{\sqrt{2}},0)$ which 
    is discussed in details in the text. Here, $M=55$, $N=3 M$ and $V_0 = 20 
    E_{r}$.}
  \label{fig:T0miPhaseSpace}
\end{figure}

\begin{figure}[t]
  \centering
  \subfigure[\: $F(\phi, \theta = \theta_{d})$]
  {\includegraphics[scale=0.40]{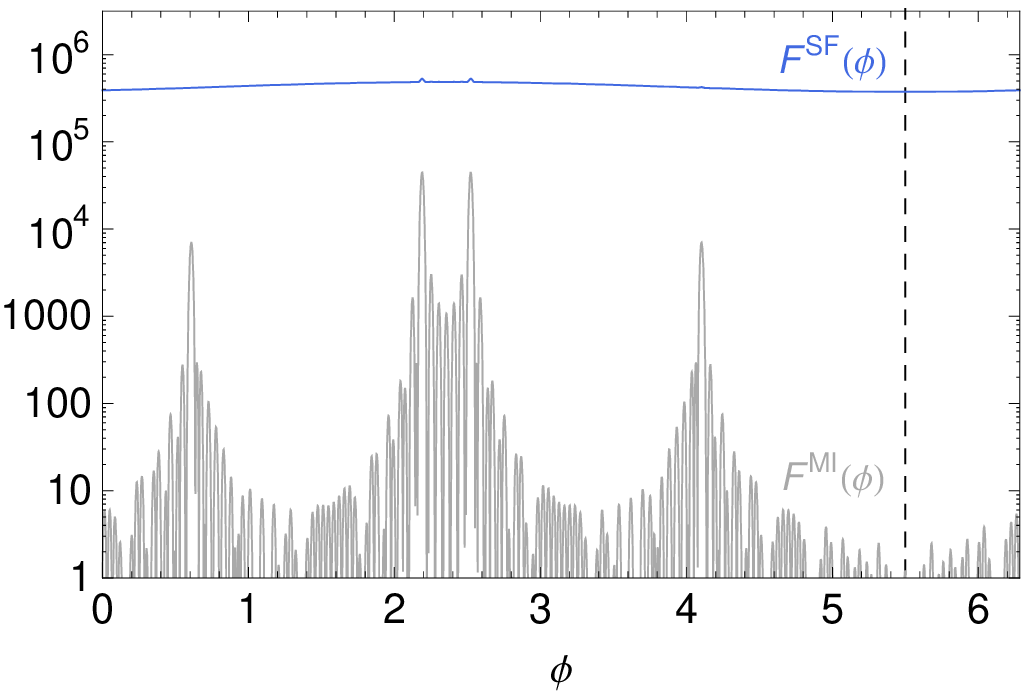}}
  \subfigure[\: $F(\phi = \phi_{d}, \theta)$]
  {\includegraphics[scale=0.40]{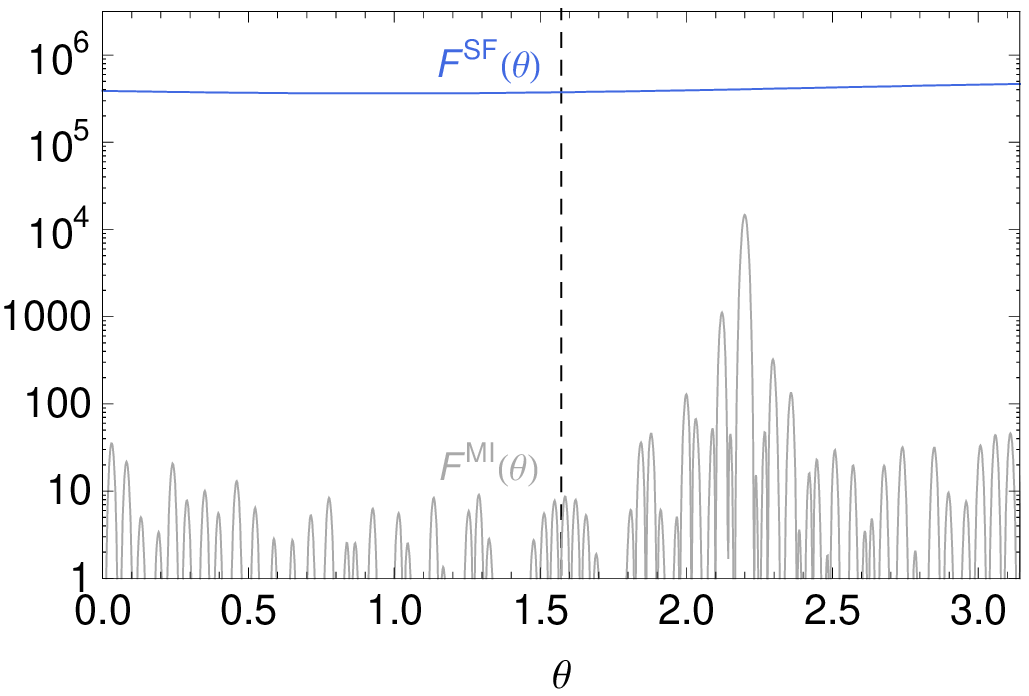}}
  \caption{(Color online) Zero-temperature $F^{SF}(\theta, \phi)$ (top blue 
    curves) and $F^{MI}(\theta, \phi)$ (bottom gray curves) with domains 
    restricted to $\phi$ (left figure) and $\theta$ (right figure). The 
    position of the photons detector $({\phi}_d, {\theta}_d)$ is indicated 
    with vertical dashed lines. Here, 
    $M=55 \!\times\! 55 \!\times\! 55$, $N= 3 M$, $\lambda_{p} / \lambda_{L}=1$ 
    and $V_0=20 E_{r}$.}
  \label{fig:T0compareMISF}
\end{figure}

In analogy to the one-dimensional case, we expect that sharp 
differences in the intensity of light scattered from the SF and 
MI phases can be observed at angles for which the classical 
component $F^{clas}(\phi, \theta)$ is negligible. In Fig.~\ref{fig:T0miPhaseSpace} 
we present the zero-temperature structure function $F(\phi,\theta)$ for MI 
phase that is equivalent to the $F^{clas}(\phi, \theta)$. Keeping in mind 
that the quantum component $F^{quant}(\phi, \theta)$ for the SF 
state is slowly varying and of order of $N$, we observe that 
$({\phi}_d, {\theta}_d)$ is indeed a promising direction for a 
measurement that can distinguish the two quantum phases. In 
Fig.~\ref{fig:T0compareMISF} we corroborate this observation by 
presenting scattering patterns for the superfluid $F^{SF}(\phi,\theta)$ 
and Mott-insulator $F^{MI}(\phi,\theta)$ phases at $T=0$. The 
plots show cross-sections of $F(\phi, \theta)$ along the planes 
of constant $\theta$ and $\phi$, respectively. Evidently, for the 
specific values of parameters we have chosen and for the assumed 
directions of the probing laser and of the detector, the difference 
of number of photons scattered from the SF and MI phases is of 
order $N \sim 10^5$ and thus should be readily measurable 
in experiment.

\begin{figure}[t]
  \centering
  \includegraphics[scale=0.4]{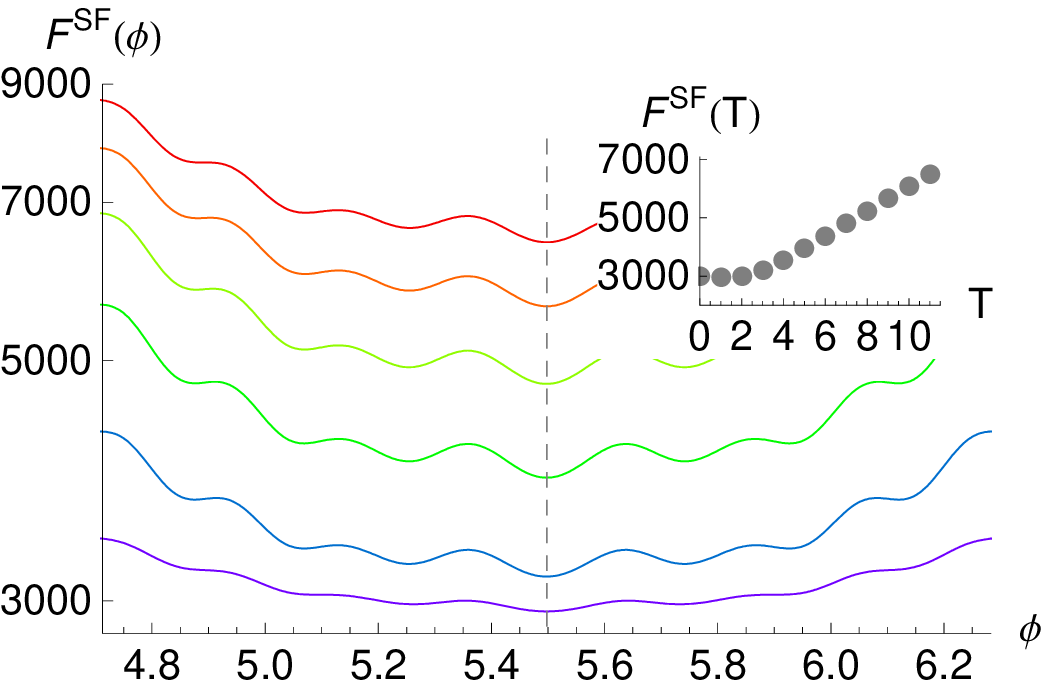}
  \includegraphics[scale=0.4]{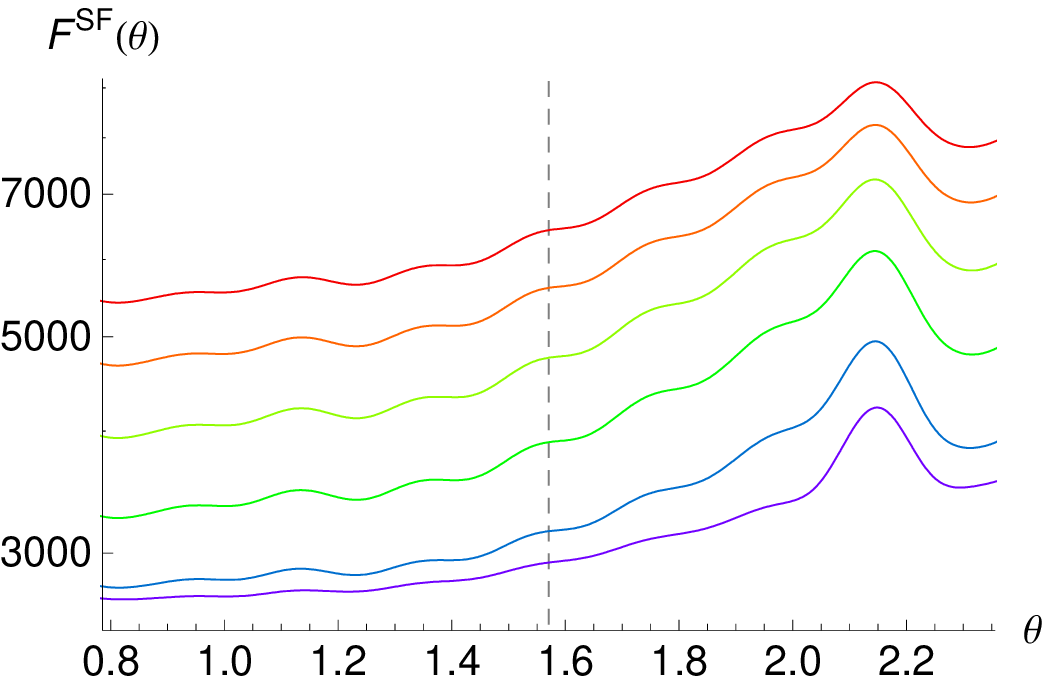} \\
  \includegraphics[scale=0.4]{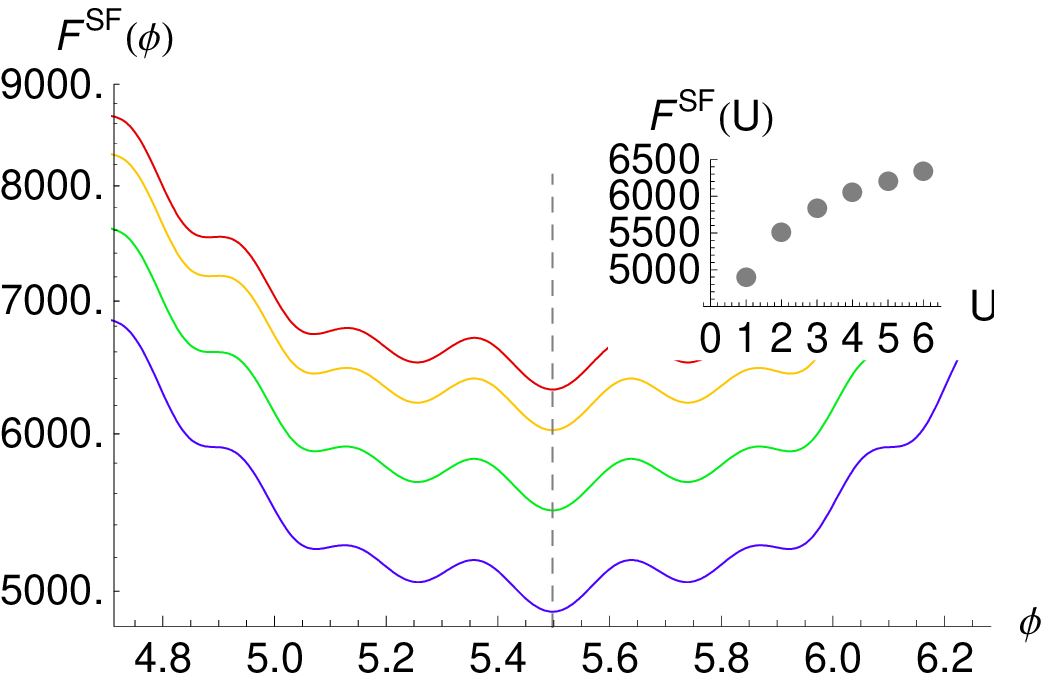}
  \includegraphics[scale=0.4]{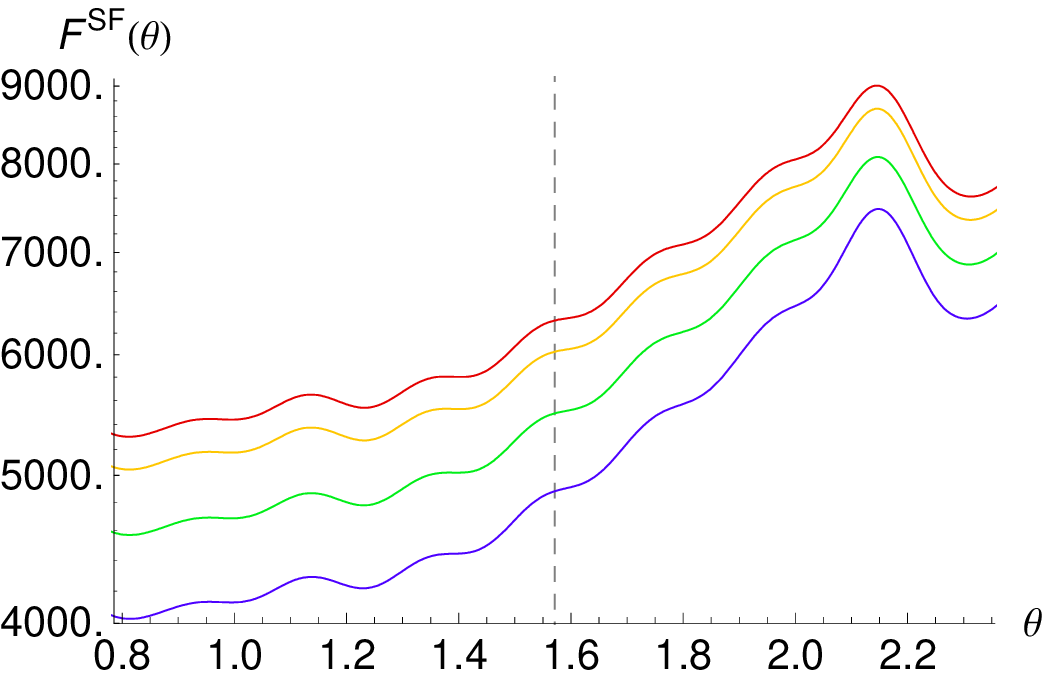}
  \caption{(Color online) Structure function $F(\phi, \theta)$ for SF phase 
    of bosons in a three-dimensional optical lattice. The preferred position of the 
    detector $({\phi}_d, {\theta}_d)$ is marked with vertical dashed lines. The plots 
    show the cross sections along the constant $\theta$ (left panels) and constant 
    $\phi$ (right panels). The calculations have been performed for $^{87}\textrm{Rb}$ 
    atoms, $\lambda_{p} = \lambda_{L} = 850\textrm{nm}$, $M=11 \!\times\! 11 \!\times\! 11$ 
    and $N= 3 M$. The upper panels show results for constant $V_0=6.80 E_r \hspace{0.07cm} 
    (U=4)$ and increasing value of temperature: $T=0, 3, 5, 7, 9, 11 $ (ordered from 
    the bottommost to the topmost curve) with $U$ and $T$ being expressed in units of 
    $J$. The bottom panels show results for constant temperature $T=10$ and increasing 
    value of trapping potential depth: $V_0 = 3.66 E_r \hspace{0.07cm} (U=1), V_0 = 5.15 
    E_r \hspace{0.07cm} (U=2), V_0 = 6.80 E_r \hspace{0.07cm} (U=4), V_0 = 7.82 E_r 
    \hspace{0.07cm} (U=6)$, ordered from the bottommost to the topmost curve. The insets 
    show the number of photons scattered in the direction of a detector 
    $({\phi}_d, {\theta}_d)$ versus temperature (top panel) and interaction strength 
    (bottom panel).}
  \label{fig:FinTsf}
\end{figure}

\begin{figure}[t]
  \centering
  \includegraphics[scale=0.4]{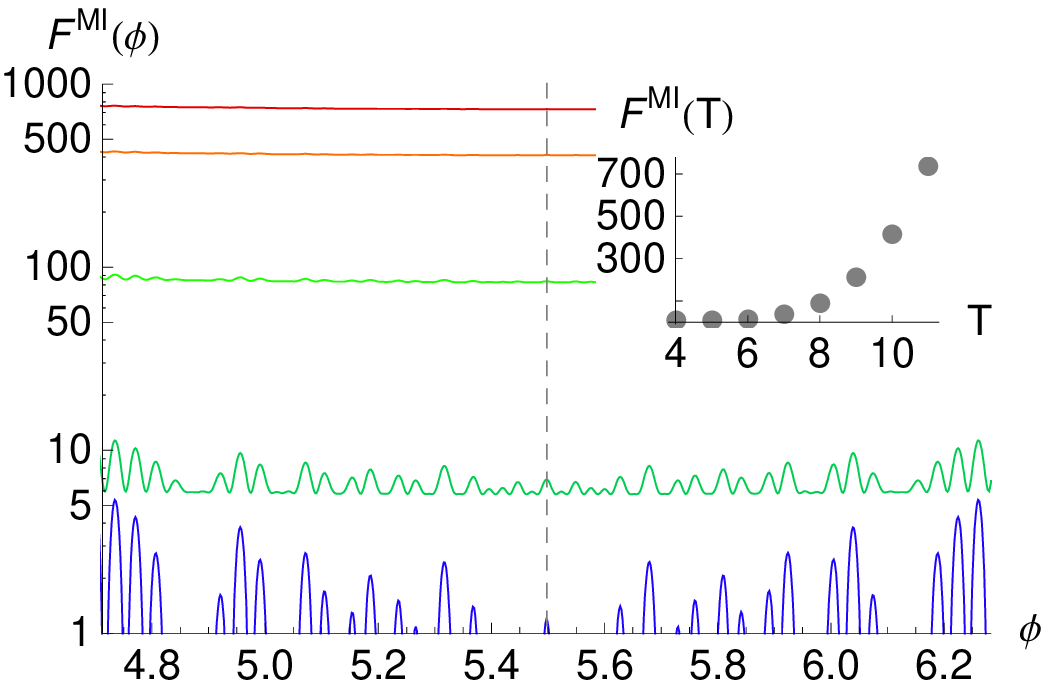}
  \includegraphics[scale=0.4]{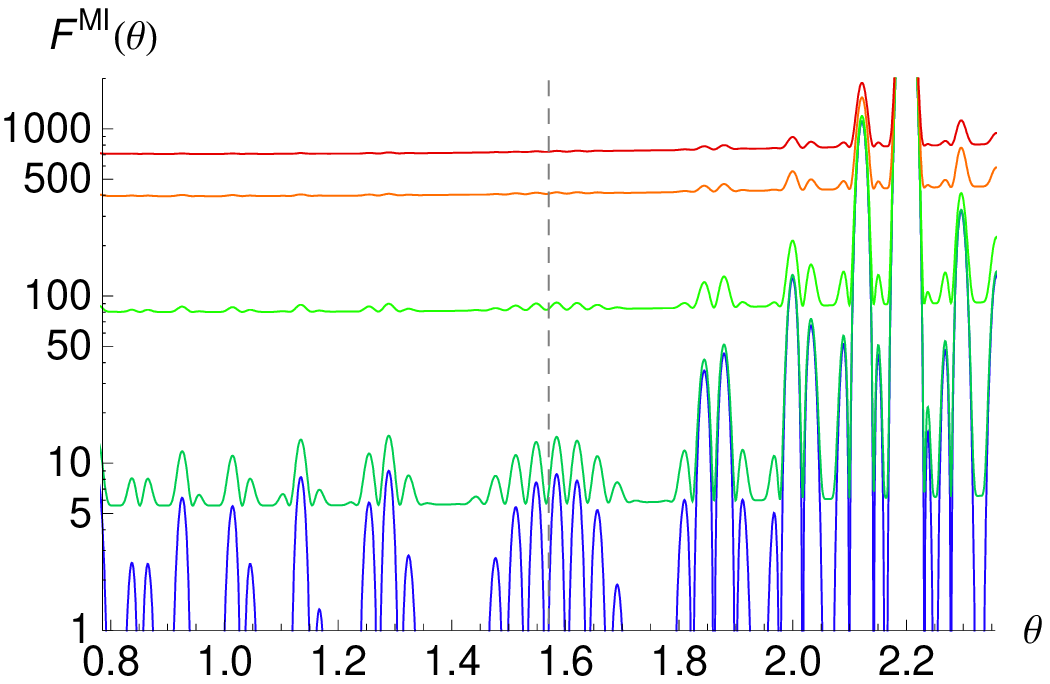}
  \caption{(Color online) Structure function $F(\phi, \theta)$ for MI phase of 
    bosons in a three-dimensional optical lattice. The preferred position of the 
    detector $({\phi}_d, {\theta}_d)$ is marked with vertical dashed lines. The 
    plots show the cross sections along the constant $\theta$ (left panels) and 
    constant $\phi$ (right panels). The calculations have been performed for 
    $^{87}\textrm{Rb}$ atoms, $\lambda_{p} = \lambda_{L} = 850\textrm{nm}$, 
    $M=55 \!\times\! 55 \!\times\! 55$, $N= 3 M$, $V_0 = 18.3 E_r \; (U=128)$, 
    $\mu = 320$ and temperatures: $T=0, 6, 8, 10, 11$ (ordered from the bottommost 
    to the topmost curve), with $U,T$ and $\mu$ being expressed in units of $J$. 
    The inset shows the number of photons scattered in the direction of a detector 
    $({\phi}_d, {\theta}_d)$ versus temperature.}
  \label{fig:FinTmi}
\end{figure}

We turn now to the thermal effects and their influence on the angular 
distribution of scattered photons. Analyzing Fig.~\ref{fig:FinTsf} and 
Fig.~\ref{fig:FinTmi} we observe the isotropic and monotonic growth of 
the intensity of scattered light with temperature, for both SF and MI 
phases. In the case of SF phase, this behavior can be explained on the 
grounds of the Eq.~(\ref{eqn:FdeepLatt}) rewritten in the momentum 
representation by means of transformation (\ref{eqn:anniFour})-(\ref{eqn:creatFour}).
In particular, if we disregard anomalous averages while calculating 
expectation values of the form 
$\aver{a^{\dagger}_{\mbf{k}_1} a_{\mbf{k}_2} a^{\dagger}_{\mbf{k}_3} a_{\mbf{k}_4} }$, 
i.e. if we perform the approximation
\begin{align}
  \aver{a^{\dagger}_{\mbf{k}_1} a_{\mbf{k}_2} a^{\dagger}_{\mbf{k}_3} a_{\mbf{k}_4} } & \approx  
  \delta_{\mbf{k}_1, \mbf{k}_2} \delta_{\mbf{k}_3, \mbf{k}_4} \aver{a^{\dagger}_{\mbf{k}_1} 
    a_{\mbf{k}_1} a^{\dagger}_{\mbf{k}_2} a_{\mbf{k}_2}} \nonumber \\
  & \phantom{\simeq} + \delta_{\mbf{k}_1, \mbf{k}_4} \delta_{\mbf{k}_2, \mbf{k}_3} 
  \aver{a^{\dagger}_{\mbf{k}_1} a_{\mbf{k}_2} a^{\dagger}_{\mbf{k}_2} a_{\mbf{k}_1}},
\end{align}
Eq.~(\ref{eqn:FdeepLatt}) can be rewritten as:
\begin{align}
  F(\mbf{q}) =& \frac{1}{M^2} \left| f_{\textbf{0},\textbf{0}} (\mbf{q}) \right|^2 \! 
  \Biggl[ N(N-1) \left| \sum\limits_{\mbf{m}} e^{\imath \mbf{q} \mbf{r}_{\mbf{m}}} \right|^2 
  \nonumber \\
  &+ \sum_{\mbf{k} \neq \mbf{k}'} \aver{n_{\mbf{k}} n_{\mbf{k}'}} \left| 
    \sum\limits_{\mbf{m}} e^{\imath \left(- \mbf{k} + \mbf{k}' + \mbf{q} \right) 
      \mbf{r}_{\mbf{m}}} \right|^2 \Biggr] \nonumber \\
  &+ N  \left| f_{\textbf{0},\textbf{0}} (\mbf{q}) \right|^2.
  \label{eqn:FMomSpace}
\end{align}
When temperature increases, a number of particles occupying excited modes 
grows (see Fig.~\ref{fig:SFcomboStatistics}), causing an increase in 
correlations terms $\aver{n_{\mbf{k}} n_{\mbf{k}'}}$. In consequence, the 
total intensity of the scattered light $F^{SF}(T)$ increases monotonically 
with temperature in any direction of measurement. We note that some 
correlations terms $\aver{n_{\mbf{k}} n_{\mbf{k}'}}$ start to decrease 
above some characteristic temperature (see Fig.~\ref{fig:SFcomboStatistics}). 
However this does not influence the total structure function $F^{SF}(T)$ that 
grows monotonically with $T$.

Similarly, an increase in the interaction strength (an increase in the lattice 
potential depth, equivalently) results in larger population of excited modes, 
partially due to increase in the quantum depletion. This behavior leads to a 
growth of the correlation terms in Eq.~(\ref{eqn:FMomSpace}) and again 
to the monotonic increase in the full function $F^{SF}(U)$.

In the case of MI phase, presented in Fig.~\ref{fig:FinTmi}, the increase in 
the number of scattered photons $F^{MI}(T)$ is fully determined by a 
temperature-driven growth of single-site fluctuations that have been 
presented in Fig.~\ref{fig:statsMI}.

\section{Summary and conclusions}
\label{sec:conclusions}

We have investigated the scattering of a weak and far-detuned laser 
light from a system of ultracold bosons in an optical lattice. We 
have shown that the light scattering can be used as a probe of the 
on-site quantum statistics, in particular fluctuations and correlations. 
Calculating the statistics for the superfluid and Mott-insulator phases at 
finite temperatures, we have determined the angular distributions of the 
mean number of the scattered photons. The profiles of the scattered light 
are fully determined by the on-site particle number fluctuations and 
correlations and thus allow for an experimental verification of the 
present theoretical models describing the statistics in ultracold gases. 
For the 3D optical lattice we have determined the optimal geometry at 
which the contribution from the Bragg scattering pattern is minimized. 
We have shown that even at some non-optimal configurations, which can 
be more accessible from the experimental point of view, this contribution 
is sufficiently small and allows one to measure the effects of quantum 
statistics. Our main conclusion is that by careful choice of the measurement 
geometry one can distinguish between different phases, even at finite 
temperatures, and observe the effects of the finite temperature statistics 
of a Bose gas.

\begin{acknowledgments}
  The authors acknowledge support of the Polish Government Research Grants 
  for years 2007-2009 (K. \L ., Z. I.) and  for years 2007-2010 (M. T.).
\end{acknowledgments}

\appendix
\section{Corrections for weak lattice potentials due to the nonlocal Franck-Condon coefficients}
\label{app:nonlocal}

By applying the local approximation in the derivation of Eq.~\eqref{eqn:FdeepLatt} 
we have neglected contributions from the nonlocal Franck-Condon coefficients. 
Here, we calculate the leading contribution from the neglected nearest-neighbor terms. 

In the case of the MI phase the expression is proportional to $N$:
\begin{equation}
  \Delta F^{MI}(\mbf{q}) = \left( |f_{1,0}(\mbf{q})|^{2} + |f_{\textrm{-}1,0}
    (\mbf{q})|^{2} \right) \left(1+n\right) N.
  \label{eqn:poprawkaMoottaa}
\end{equation}
Although it scales the same as the difference between MI and SF phase, 
the coefficients $f_{1,0}(\mbf{q})$ and $f_{\textrm{-}1,0}(\mbf{q})$ rapidly 
tend to zero with the increasing lattice depth.

Similarly, the nearest neighbor correction for superfluid state reads
\begin{align}
  \Delta F^{SF}(\mbf{q}) & = \Big(|f_{1,0}(\mbf{q})|^2 + |f_{\textrm{-}1,0}(\mbf{q})|^2 
  \nonumber\\
  & \phantom{=} \,  +2 \textrm{Re}\!\left\{f_{1,0}^{\ast}(\mbf{q}) \,f_{\textrm{-}1,0}
    (\mbf{q}) \right\} \Big) \nonumber \\
  & \phantom{=} \times \left[N + n^2 \left(1-\frac{1}{N}\right)\left| 
      \sum\limits_{n} e^{\imath \mbf{q} \mbf{r}_{n}} \right|^2 \right].
  \label{eqn:poprawkanadcieklego}
\end{align}
A brief estimate leads to
\begin{equation*}
  \Delta F^{SF}(\mbf{q}) \approx \left(|f_{1,0}(\mbf{q})|^2 
    + |f_{\textrm{-}1,0}(\mbf{q})|^2 \right) N,
\end{equation*}
that, again, contains small coefficients $f_{1,0}(\mbf{q})$ and $f_{\textrm{-}1,0}(\mbf{q})$ 
rapidly decreasing with the lattice potential depth.

\section{Optimization of positions of a probing laser and a detector in the 3D case}
\label{app:optimization}

In this appendix we derive the condition for optimal configuration of 
the probing light and of the photon detector, which lead to the minimal 
contribution from the classical amplitude of the scattered light. We start 
with the classical part of the structure function $F(\mbf{q})$, defined in 
\eqref{eqn:FClasDef}
\begin{align}
  F^{clas}(\mbf{q}) & = n^2 \left| f_{\textbf{0},\textbf{0}} (\mbf{q}) \right|^2 \left| 
    \sum\limits_{\mbf{m}} e^{\imath \mbf{q} \mbf{r}_{\mbf{m}}} \right|^{2}.
\end{align}
The label $\mbf{m}$, enumerates lattice sites: $r_\mbf{m} = d (\hat{\mbf{x}} m_x + 
\hat{\mbf{y}} m_y + \hat{\mbf{z}} m_z)$, with integer $m_x,m_y,m_z$. For a simple 
cubic lattice the summation can be easily performed
\begin{align}
  F^{clas}(\mbf{q}) = n^2 \left| f_{\textbf{0},\textbf{0}} (\mbf{q}) \right|^2 
  \prod_{i = x,y,z} 
  \frac{\sin^2\left(\frac M 2 q_i d \right)}{\sin^2\left(\frac12 q_i d \right)}.
  \label{eqn:Fcl1}
\end{align}
For the rest of the derivation we introduce a convenient parametrization 
of the vectors $\mbf{k}_L = k_L (\alpha_x,\alpha_y,\alpha_z)$, $\mbf{k} = k_L 
(\beta_x,\beta_y,\beta_z)$, and $\mbf{q} = k_L (\eta_x,\eta_y,\eta_z)$ describing 
the momenta of the incoming and scattered photons, and the momentum transfer, 
respectively. The dimensionless numbers $\eta_i$, $\alpha_i$, and $\beta_i$ satisfy:
$|\eta_i|<2$, $|\alpha_i|<1$, and $|\beta_i|<1$ for $i = x,y,z$. Expressing 
the translation vector of the lattice and the wave vector of the laser in 
terms of the wavelengths: $d = \lambda_{p}/2$ and $k_L = 2 \pi/\lambda_L$, 
we rewrite Eq.~\eqref{eqn:Fcl1} in the following way
\begin{align}
  F^{clas}(\mbf{q}) = n^2 \left| f_{\textbf{0},\textbf{0}} (\mbf{q}) \right|^2 
  \prod_{i = x,y,z} \frac{\sin^2\left(M \frac{\pi}{2} \eta_i \frac{\lambda_{p}}
      {\lambda_L} \right)}{\sin^2\left(\frac{\pi}{2} \eta_i 
      \frac{\lambda_{p}}{\lambda_L} \right)}.
  \label{eqn:Fcl2}
\end{align}
Typically, the angular dependence of the Franck-Condon factor 
$\left| f_{\textbf{0},\textbf{0}} (\mbf{q}) \right|^2$ is rather weak, which 
follows from the fact that the characteristic size of a single lattice site, 
given by a harmonic oscillator length associated with the potential well, is 
much smaller than the wavelength of the probing laser. In such conditions the 
scattering due to $\left| f_{\textbf{0},\textbf{0}} (\mbf{q}) \right|^2$ is 
almost isotropic, and most of the angular dependence is determined by the interference 
term characteristic for the Bragg scattering. The function $\sin^2(M x)/\sin^2(x)$ 
generating the interference pattern, takes the maxima at $x = n \pi$, while the minimal 
amplitude of oscillations occurs in the middle between two neighboring maxima: 
$x= \pi(n + \frac12)$. In fact, the latter determines the desired condition for 
the measurement with the minimal contribution from the classical component: 
$\eta_i \frac{\lambda_{p}}{\lambda_L} = 1 + 2 n_i$, where $n_i$ are integers, 
and $i = x,y,z$. For simplicity we further consider only the simplest case 
$\lambda_{p} = \lambda_L$. Since $|\eta_i|<2$, the only possibility is 
$\eta_i = \pm 1$, which leads to the following three conditions:
\begin{align}
  \label{eq:cond1}
  \beta_j - \alpha_j = \pm 1, \qquad \textrm{for }j=x,y,z.
\end{align}
The other two conditions are given by the conservation of the momenta of 
the scattered photons: $|\mbf{k}| = |\mbf{k_L}| = k_L$, which results in
\begin{equation}
  \label{eq:cond2}
  |\alpha_x|^2 + |\alpha_y|^2 + |\alpha_z|^2 = |\beta_x|^2 + |\beta_y|^2 
  + |\beta_z|^2 = 1.
\end{equation}
By combining \eqref{eq:cond1} and \eqref{eq:cond2}, we obtain the following 
two equations determining the coordinates of $\mbf{k}$ and $\mbf{k}_L$:
\begin{align}
  \label{eq:cond3}
  |\alpha_x|^2 + |\alpha_y|^2 + |\alpha_z|^2 = 1, \\
  \label{eq:cond4}
  |\alpha_x \pm 1|^2 + |\alpha_y \pm 1|^2 + |\alpha_z \pm 1|^2 = 1.
\end{align}
Readily, there are infinitely many solutions of the two above equations. All 
of them lie on a circle that is a common part of two spheres in the three-dimensional 
space. One of the possible solutions is given by the set of numbers: 
$\mbf{k}_L = k_L \left(6-\sqrt{6},6-\sqrt{6},6+2\sqrt{6}\right)/12$ and 
$\mbf{k} = k_L \left(-6-\sqrt{6},-6-\sqrt{6},2\sqrt{6}-6\right)/12$.

Finally, we note that for the optimal geometry determined by Eqs.~\eqref{eq:cond3} 
and \eqref{eq:cond4}, an average of the classical component $F^{clas}(\mbf{q})$ over 
a finite solid angle containing several interference peaks results in the 
three-dimensional analog of formula \eqref{eqn:FMI0rdAvg}
\begin{align}
  \overline{F^{clas}(\mbf{q})} = \left| f_{\mbf{0},\mbf{0}} (\mbf{q}) \right|^2 
  \frac{n^2}{8}.
  \label{eqn:FMI0rdAvg3D}
\end{align}
Here, we have applied the condition $\sin\left(\frac12 q_i d \right) =1$ that 
follows the conditions for the optimal choice of the measurement geometry. 
We stress that this result is derived for this particular geometry, and only in 
this case the sine squared factors average out to $\frac{1}{2}$ independently in 
all three directions.

\bibliographystyle{apsrev}
\bibliography{thermalMISFscatt_v3}{}

\begin{thebibliography}{39}
\bibitem[{rem({\natexlab{a}})}]{rem1}
\bibinfo{note}{We note that $S(\mbf{q})$ is frequently defined in the
  literature as the Fourier transform of density fluctuations $S(\mbf{q}) =
  N^{-1} \int \!\!d^{3}x \!\! \int \!\!d^{3}y \: e^{\imath \mbf{q} \left(
  \mbf{x} - \mbf{y} \right) }\left\langle
  \hat{\rho}(\mbf{x})\hat{\rho}(\mbf{y}) \right\rangle - \left\langle
  \hat{\rho}(\mbf{x})\right\rangle \left\langle \hat{\rho}(\mbf{y})
  \right\rangle$ \cite{Pitaevskii}}.
\end{thebibliography}
\end{document}